\pdfoutput=1
\documentclass[final]{IEEEtran}

\usepackage{amsmath,amssymb,amsfonts,mathrsfs,bm,amsthm}
\usepackage{amstext}
\usepackage{upgreek}
\usepackage{multicol}
\usepackage{indentfirst}
\usepackage{graphicx}
\usepackage{paralist}
\usepackage{multirow}
\usepackage{tabularx}
\usepackage{cite}
\usepackage{booktabs}
\usepackage{subfigure}
\usepackage{color}
\usepackage{algorithm,algorithmic}

\newtheorem{corollary}{\bf Corollary}
\newtheorem{proposition}{\bf Proposition}

\allowdisplaybreaks
\usepackage{flushend}
\begin{document}
\title{Caching at Base Stations with Heterogeneous User Demands and Spatial Locality}
\author{Dong Liu and Chenyang Yang
	\thanks{This work was supported in part by National Natural Science Foundation of China (NSFC) under Grants with No. 61731002 and 61671036. This paper was presented in part at the IEEE ICC Workshops 2018~\cite{dongws}.}
	\thanks{The authors are with School of Electronics and Information Engineering, Beihang University, Beijing, China (e-mail: \{dliu, cyyang\}@buaa.edu.cn).}
}
\maketitle

\begin{abstract}
Existing proactive caching policies are designed by assuming that all users request contents with identical activity level at uniformly-distributed or known locations, among which most of the policies are optimized by assuming that user preference is identical to content popularity. However, these assumptions are not true based on recent data analysis. In this paper, we investigate what happens without these assumptions. To this end, we establish a framework to optimize caching policy for base stations exploiting heterogeneous \emph{user preference}, \emph{activity level}, and \emph{spatial locality}. We derive success probability and  average rate of each user as utility function, respectively, and obtain the optimal caching policy maximizing a weighted sum of average utility (reflecting network performance) and minimal utility of users (reflecting user fairness). To investigate the intertwined impact of individual user request behavior on caching, we provide an algorithm to synthesize user preference from given content popularity and activity level with controlled preference similarity, and validate the algorithm with real datasets. Analysis and simulation results show that exploiting individual user behavior can improve both network performance and user fairness, and the gain increases with the skewness of spatial locality, and the heterogeneity of user preference and activity level.
\end{abstract}


\section{Introduction}
Owing to the 80-20 rule in terms of user behavior in requesting contents, caching at the wireless edge is a promising approach for supporting the ever-increasing wireless data traffic.
By caching at BSs, the traffic load of backhaul and service latency of users can be reduced, which improves network throughput, energy efficiency and user experience dramatically~\cite{bigdata,liu2016energy}. By caching at user devices, users can fetch the requested contents directly from their own storage and/or from nearby users via device-to-device (D2D) communications, which can further offload wireless traffic~\cite{Niki13,czy}.
Facing the limited storage size at wireless edge while with huge number of contents, optimizing proactive caching policy by exploiting the skewed distribution of user demands is critical in reaping the benefit of wireless edge caching~\cite{bigdata,liu2016energy,Niki13,czy}.

Most proactive caching policies are optimized based on \emph{content popularity},
under the assumptions of exactly known or completely unknown user locations.
Assuming that the location where each user sends request is known \emph{a priori} when optimizing caching policy, deterministic caching polices were proposed in~\cite{femtocachingTIT,BER,chenzhen}. A policy minimizing the average download delay was proposed in~\cite{femtocachingTIT}, which was shown with minor performance loss when the users are with unknown locations by simulation. The caching policies were respectively optimized to minimize average bit error rate  in~\cite{BER}, and to maximize the successful transmission probability for cellular network with cooperative transmission in~\cite{chenzhen}. Considering that user locations are hard to predict, user and BS locations are assumed as Poisson point process (PPP) in~\cite{Blaszczyszyn2015optimal,wen2017cache,czy,xiuhua,TMX}. A probabilistic caching policy was proposed in~\cite{Blaszczyszyn2015optimal} and then extended into multi-tier heterogeneous networks in~\cite{wen2017cache}  to maximize the success probability, and was optimized to maximize the cache-hit probability in~\cite{czy}, where the policy for every BS in the same tier is identical. A deterministic caching policy was jointly optimized with user association to maximize the supported traffic load in~\cite{xiuhua}, and coded caching policies were optimized that respectively maximize average fractional offloaded traffic and average ergodic rate in~\cite{TMX}. Both works deal with unknown user locations by deriving the probability of a user associated with a BS. A coded caching policy was optimized to minimize energy consumption in~\cite{EE}, which deals with unknown user locations by assuming known probability that a user is in the coverage of a BS.

While proactive caching is motivated by the Pareto principle for user behavior, owing to the isolation among different disciplines, the following facts regarding user behavior are largely overlooked in the literature of proactive caching: 

1) \emph{Content popularity reflects average interests of multiple users}, but cannot reflect the preference of an individual user. This is because user preferences are heterogeneous, which has been widely acknowledged in recommendation systems~\cite{ekstrand2011collaborative}.

2) \emph{Activity levels of users are heterogeneous}. As reported in~\cite{traffic,traffic2}, 80\% of the daily network traffic is generated by less than 20\% of all users. 

3) \emph{A user does not send requests in every cell with
equal probability (i.e., exhibits spatial locality), and users have different spatial distributions when sending requests}.
As reported in~\cite{traffic,traffic2,Understanding}, most mobile users periodically initiate content requests in limited number of locations with high probability. Specifically, big data analysis in~\cite{Understanding} shows that 80\% of the users only send requests for contents from less than four places, which implies that the probability that in which cell a user is located when sending requests is predictable from the request history. Hence, existing assumptions on user location (i.e., perfectly known or uniformly distributed throughout the network) are either too optimistic or too pessimistic.


User preference can be predicted via machine learning techniques such as collaborative filtering~\cite{hofmann2004latent,ekstrand2011collaborative} and deep learning~\cite{DL}, which has been leveraged in wireless networks recently~\cite{bigdata,zhang2016clustered,CBQ,liujuan,Preference}. An
immediate way to employ user preferences for caching is to aggregate them into
local content popularity of a cell as proposed in~\cite{bigdata}, where the preferences are learned at the core network of a mobile network operator (MNO) by monitoring and analyzing historical traffic.
While caching at BSs should consider the demand statistics of all users in a cell, the performance may differ by exploiting the coarse-grained user behavior (i.e., content popularity) and fine-grained behavior (i.e., user preference and spatial locality). The gain from caching at the wireless edge could be further improved if the caching policies are optimized directly with individual user preference as in~\cite{zhang2016clustered,CBQ,Preference} for D2D communications and in~\cite{liujuan} for fog radio access networks.
To evaluate the performance of proposed caching policies, user preferences were assumed as Zipf distributions with different ranks in~\cite{zhang2016clustered,liujuan} without validation, and were synthesized in~\cite{Preference} with a hierarchical parametric model proposed in~\cite{Individual} based on a real dataset.


Heterogeneous spatial locality, user preference and activity level introduce new challenges and possible benefits into wireless edge caching. Existing framework based on fixed user locations~\cite{femtocachingTIT,chenzhen,BER,liujuan} or uniformly distributed user and BS locations modeled by homogeneous PPP~\cite{Blaszczyszyn2015optimal,wen2017cache,czy,xiuhua,TMX} cannot capture the spatial locality. When each user with individual preference sends file request in some cells with high probability, the user locations  are no longer  independent with BS locations, and the users are no longer equivalent among each other as implied by the PPP model. Although more complex model, such as Poisson Cluster Process~\cite{PCP}, can capture the coupling between BS and user locations, the non-equivalence among users due to the heterogeneity in both user preference and spatial locality still makes the model not applicable.
Moreover, since users are no longer statistically equivalent to each other, maximizing the network average performance, e.g., the success probability or average rate of a randomly chosen user, cannot let every user benefit from caching. The \emph{caching interests} of users may conflict, e.g., a user may prefer a BS to cache one file while the other user may prefer the BS to cache another file. Hence, caching policy will affect the fairness among users. This indicates that caching can bring a new dimension to addressing the user fairness issue in application level, which differs from the traditional way of improving fairness with radio resource allocation based on channel information.


In this paper, we investigate when and how spatial locality and preference heterogeneity of users impact caching.
Since no existing optimization frameworks can be applied for this purpose, we establish a new framework, where
BS locations are fixed (rather than uniformly distributed), which are true in practical networks, and
users are non-uniformly located among different cells. We improve both network performance and user fairness, taking max-min fairness~\cite{maxmin} as an example. We consider success probability or average achievable rate as \emph{user utility} and optimize caching policy to maximize a weighted sum of average user utility and minimal user utility.

The major contributions of this paper are summarized as follows:
\begin{itemize}
	\item  Instead of assuming user location as perfectly known or completely unknown, we assume that the probabilities of each user sending requests in different cells are known, which can capture the spatial locality of individual user. We consider a general probabilistic caching policy, where the BSs can cache files based on different probability distribution to accommodate the user preference heterogeneity and the user spatial locality.
	\item Different from existing literature only maximizing the average performance, we consider both network performance and user fairness in optimization. We show that the optimization problem is equivalent to a non-convex signomial programming problem. We solve the problem efficiently by successively solving a series of convex problems, and analyze the behavior of the policy under special cases. Our results show that exploiting individual user behavior can improve both metrics remarkably, whose gain is large when user preferences are less similar, spatial locality is strong, and user activity level is heterogeneous.
	\item We examine the assumptions on individual user behavior in requesting contents by analyzing two real datasets~\cite{MSD,Lastfm}. We provide an algorithm to synthesize user preference from the data generated with given content popularity, which can be used for caching policy evaluation with flexibly controlled user behavior statistics and is validated by  real datasets.
\end{itemize}

The rest of the paper is organized as follows. In Section~II,
we present the system model and define user behaviors and show their relations. Section III derives the success probability and average rate of users, and optimizes the
caching policy exploiting user preference and spatial locality. Section~IV analyzes user behavior from real datasets and proposes a user preference synthesization algorithm. Simulations and conclusions are provided in Sections V and VI, respectively.

\section{System Model and User Behavior in Requesting Contents}
\begin{figure}[!htb]
	\centering	
	\subfigure[Random BS locations, $K = 2$]{
		\label{fig:layout_general} 
		\includegraphics[height=0.23\textwidth]{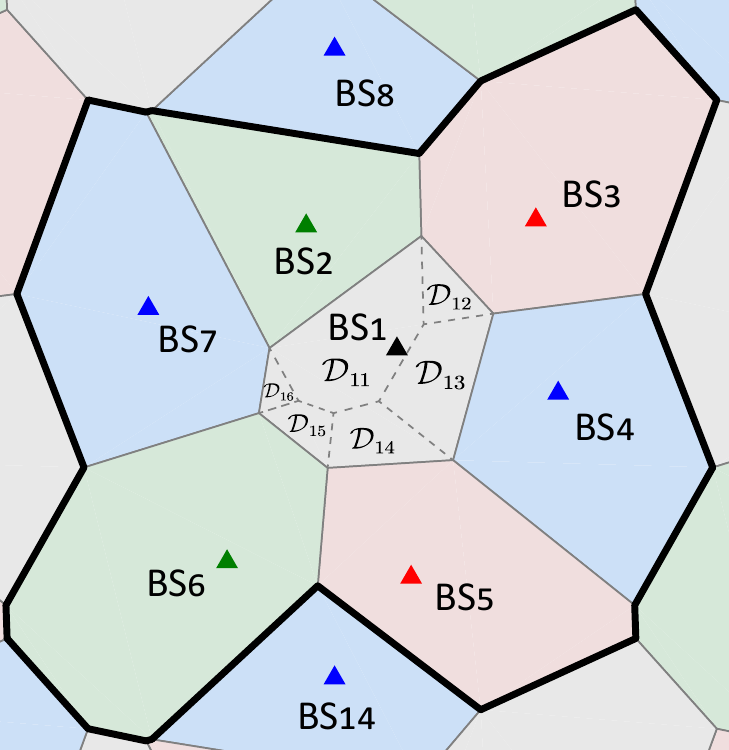}}
	\subfigure[Hexagonal cells, $K=3$]{
		\label{fig:layout} 
		\includegraphics[height=0.23\textwidth]{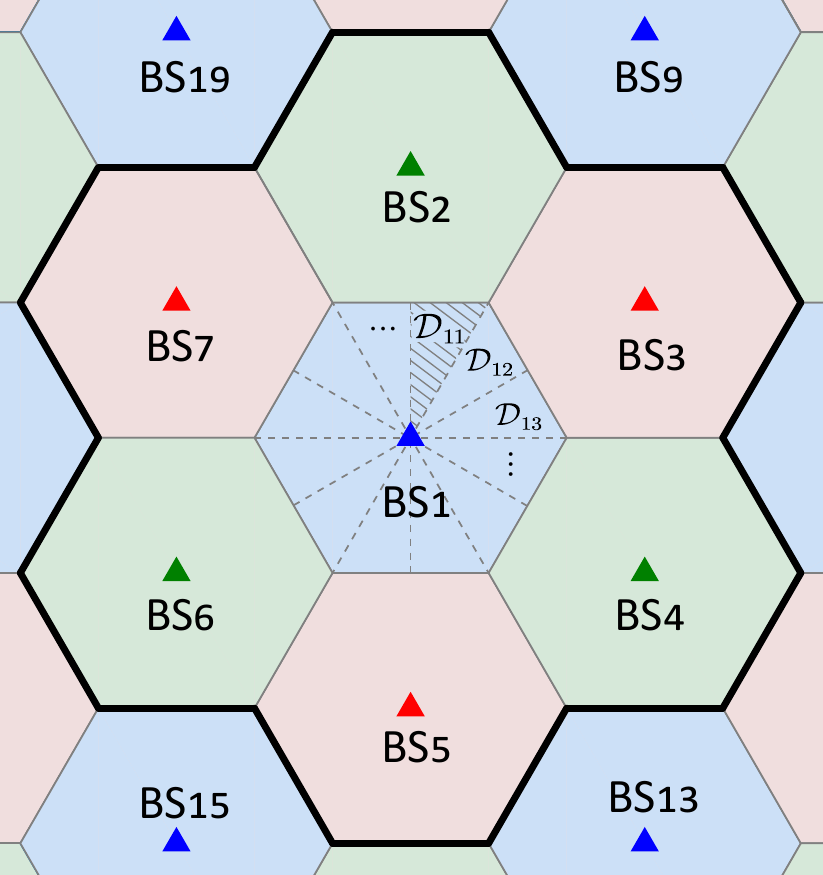}}
	\caption{Example layout of cache-enabled networks with irregular and regular cells. The considered region is surrounded by bold line. In the two examples, $N_b = 7$. Without loss of generality, the origin is set at BS$_1$, i.e., $\mathbf y_1 = (0,0)$.}
	\label{fig:network}
\end{figure}

Consider a cache-enabled wireless network, where each BS is equipped with a cache and connected to the core network via limited-capacity backhaul. The considered region contains $N_u$ users and $N_b$ BSs as shown in Fig. \ref{fig:network}, where two example layouts are provided respectively with Voronoi tessellation cell boundaries and with hexagonal cells.

\subsection{Caching Policy and User Association}
We consider a general probabilistic caching policy to accommodate heterogeneous user preference and spatial locality by allowing each BS to cache files with different probability distribution. Each BS can cache at most $N_c$ files from a content library consisting of $N_f$ equal-sized files that all the users in the considered region may request. Denote $c_{fb}$ $(0\leq c_{fb} \leq 1)$ as the probability that the $b$th BS caches the $f$th file. To realize the probabilistic caching policy  for fixed BS locations, each BS can determine which specific files should be cached  based on $\{c_{fb}\}_{b=1,\cdots,N_b, f=1,\cdots,N_f}$ by the method in~\cite{Blaszczyszyn2015optimal} periodically (e.g., in every few hours to reduce the overhead for content replacement). Considering that user preference changes much slower than traffic load, the caching policy can be optimized and updated during off-peak time.
When $c_{f1} = \cdots = c_{fN_b}$, the caching policy is identical for every BS, as existing caching policies in homogeneous networks~\cite{Blaszczyszyn2015optimal}, or as exsiting caching polices for BSs in the same tier of heterogeneous networks~\cite{wen2017cache}. When $c_{fb}\in \{0,1\}$, it degenerates into deterministic caching policy.

Since the coverage of BSs could be overlapped, to increase the cache-hit probability, each user is allowed to associate with one of the $K$-nearest neighbor BSs (called \emph{neighboring BS set}) to download the requested file from the BS's cache. Then, each (irregular or regular) cell can be divided into several small regions formed by the $K$-nearest neighbor Voronoi tessellation~\cite{knn} (shown by dashed lines in Fig.~\ref{fig:network}), so that  for a user in each small region the neighboring BS set  is fixed. For example, when a user is located in $\mathcal{D}_{11}$ of Fig.~\ref{fig:layout_general} with $K = 2$, the nearest BS and the second nearest BS are BS$_1$ and BS$_2$, respectively, where $\mathcal{D}_{ij}$ denotes the $j$th small region of the $i$th cell.
The nearest BS is called the \emph{local BS} of the user.

Since caching is more beneficial for networks with stringent-capacity backhaul~\cite{Niki13}, we assume that backhaul is the bottleneck for content delivery, i.e., a user can achieve higher data rate when downloading from the cache than from the backhaul. Therefore, if the requested file is cached in the neighboring BS set, the user will associate with the nearest BS\footnote{For mathematical tractability, we do not consider shadowing, which will not change the main	trends of the performance.} that caches the requested file and download from the cache. Otherwise, the user will associate with the local BS and fetch the file via backhaul. To avoid strong inter-cell interference inside the neighboring BS sets, especially the interference generated from the local BS to a user when the user downloads file from other BSs~\cite{liu2016energy}, the BSs within the neighboring BS set use different frequency resource.

Then, the receive signal-to-interference-plus-noise ratio (SINR) of the $u$th user when it is located at $\mathbf{x}_u$ and downloads from the $b$th BS is given by
\begin{equation}
\gamma_{ub}(\mathbf{x}_u) = \frac{Ph_{ub}r_{ub}^{-\alpha}}{\sum_{b'\in \Phi_b, b' \neq b} P h_{ub'} r_{ub'}^{-\alpha} + \sigma^2 } \triangleq \frac{S_{ub}}{I_{ub} + \frac{\sigma^2}{P}}
\end{equation}
where $P$ is the transmit power of BS, $h_{ub}$ and $r_{ub} = || \mathbf{x}_u - \mathbf{y}_b ||$ are respectively the channel power and Euclidean distance between the $u$th user and the $b$th BS, $\mathbf{x}_u = (x_{u1}, x_{u2})$ and $\mathbf{y}_{b} = (y_{b1}, y_{b2})$ are respectively the coordinates of the $u$th user and the $b$th BS, $\alpha$ is the pathloss exponent, $\Phi_b$ denotes the BS set that shares the same frequency with the $b$th BS, and $\sigma^2$ is the noise power. We consider Rayleigh fading and hence $h_{ub}$ follows exponential distribution with unit mean.

\subsection{User Behavior in Requesting Contents}
\emph{Spatial Locality} of a user is captured by its location probability distribution, denoted as $\mathbf{a}_u =[a_{u1},\cdots, a_{u_{N_b}}]$ for the $u$th user, where $a_{ui}$ is the probability that the user is located in the $i$th cell when initiating a file request. The user location probability matrix is denoted by $\mathbf{A} = [\mathbf a_1^T, \cdots, \mathbf{a}_{N_u}^T]^T$. Since proactive caching policy is optimized during off-peak time, which might be hours in advance to the time of delivering content, the exact location (e.g., $\mathbf{x}_u$) where a mobile user will send a request is hard to predict. To reflect the predictability of $\mathbf A$ and the uncertainty on the exact location, we assume that  $\mathbf A$ is known, but a user is uniformly distributed in a cell if the user sends request in the cell. Our work can be easily extended if fine-grained prediction for user location (e.g., the probability that the $u$th user is located in $\mathbf{x}_u$) can be predicted.

\emph{Global Content Popularity} is the probability distribution of the file requests in the considered region, denoted as $\mathbf{p} = [p_1, \cdots, p_{N_f}]$, where $p_f$ is the probability that a file requested by the users in the $N_b$ cells is the $f$th file.

\emph{Local Content Popularity} is the probability distribution of the requests in one cell, denoted as $\mathbf{p}_i = [p_{1i}, \cdots, p_{N_fi}]$ for the $i$th cell, where $p_{fi}$ is the probability that a file requested by the users in the $i$th cell is the $f$th file.

\emph{User Preference} of a user is the probability distribution of the requests from the user, denoted  as $\mathbf{q}_u = [q_{u1}, \cdots, q_{uN_f}]$ for the $u$th user, where $q_{uf} \in [0,1]$ is the probability that a file requested by the $u$th user is the $f$th file, and let $\mathbf{Q} = [\mathbf{q}_1^T, \cdots, \mathbf q_{N_u}^T]^T$ denote the user preference matrix.

\emph{User Activity Level Heterogeneity} is captured by a distribution denoted as $\mathbf{v} = [v_1, \cdots, v_{N_u}]$, where $v_u$ is the probability that a file request in the $N_b$ cells is sent from the $u$th user.

Based on the law of total probability, the relation between the global content popularity and user preference for the $f$th file can be expressed as
\begin{equation}
p_{f}  = \sum_{u=1}^{N_u} v_uq_{uf} = \mathbb{E}_u [q_{uf}]  \label{eqn:relation}
\end{equation}
where $\mathbb{E}_u$ denotes the expectation with respect to $u$, i.e., the requesting user. This relation shows that global content popularity is the average of user preferences in a region.

Similarly, the relation between the local content popularity in the $i$th cell and user preference  for the $f$th file can be expressed as
\begin{equation}
p_{fi} = \frac{\sum_{u = 1}^{N_u}a_{ui} v_u q_{uf}}{\sum_{f = 1}^{N_f}\sum_{u = 1}^{N_u}a_{ui} v_u q_{uf}} = \frac{\sum_{u = 1}^{N_u}a_{ui}v_u q_{uf}}{\sum_{u = 1}^{N_u}a_{ui}v_u} \label{eqn:local}
\end{equation}

Either when the shape of probability distribution $\mathbf{q}_u$ differs from that of  $\mathbf{q}_m$, or when the rankings of the elements in $\mathbf{q}_u$ and $\mathbf{q}_m$ differ, the two users have different preferences. To reflect the relation between the preferences of two users, we consider cosine similarity frequently used in collaborative filtering~\cite{ekstrand2011collaborative}, defined as $\cos(\mathbf{q}_u, \mathbf{q}_{m}) \triangleq \frac{\mathbf{q}_u\mathbf{q}_m^T}{||\mathbf{q}_u||\cdot||\mathbf{q}_m||}$.
 To use one parameter to characterize the heterogeneity of user preference in a region, we consider average similarity, which is the cosine similarity averaged over all the two-user pairs,~i.e.,
\begin{equation}
{\rm sim}(\mathbf{Q}) \triangleq \frac{2}{N_u(N_u - 1)}\sum_{u = 1}^{N_u - 1}\sum_{m = u + 1}^{N_u} \cos(\mathbf{q}_u, \mathbf{q}_{m}) \label{eqn:avecos}
\end{equation}

In practice, user preference $\mathbf{Q}$ and activity level $\mathbf{v}$ can be learned by implicit feedback collaborative filtering techniques such as probabilistic latent semantic analysis~\cite{hofmann2004latent,CBQ} or matrix factorization\cite{hu2008collaborative,bigdata}, and user location probability $\mathbf{A}$ can also be learned by machine learning, all at the service gateway of MNO by analyzing historical requests. They are assumed perfect in this work, since our focus is to find when exploiting individual user behavior is beneficial.

From \eqref{eqn:relation} and \eqref{eqn:local}, we can obtain the following observation.

{\bf Observation:}
\emph{
\begin{enumerate}
	\item If user preferences are homogeneous  (i.e., $q_{1f} = \cdots = q_{N_uf}$, $\forall f$), we have $p_{f} = p_{fi} = q_{uf}$, i.e., there is no difference between global and local content popularity as well as user preference, no matter if the demands of each user exhibits spatial locality.
	\item If each user is not with spatial locality (i.e., $a_{u1} = \cdots = a_{uN_b}$), we have $p_{f} = p_{fi}$, no matter if user preferences are identical.
	\item When user preferences are heterogeneous and user location distribution is non-uniform, global, local content popularity and user preference are different.
\end{enumerate}	
}

\section{Caching Policy Optimization With Individual User Behavior }
We consider two widely adopted metrics for content delivery, average achievable rate and success probability.\footnote{We can also consider successful transmission probability, i.e., the probability that a user can download the requested file from cache with achievable rate larger than a threshold. } The success probability   is defined as the probability that a user can download the requested file from cache with receive SINR larger than a threshold $\gamma_0$~\cite{wen2017cache,Blaszczyszyn2015optimal}.

In this section, we first derive the success probability and  average achievable rate of each user considering spatial locally and heterogeneous user behavior. Then, we establish a framework to optimize caching policy that improves both network performance and user fairness. Since the optimal policy is not with closed-form expression, we demonstrate its behavior analytically in special cases and numerically with toy examples.

Let $k(\mathbf{x}_u)$ denote the index of the $k$th nearest BS of the $u$th user located at $\mathbf{x}_u$. Then, $1(\mathbf{x}_u)$ denotes the index of the local BS (i.e., the  nearest BS) of the $u$th user. Based on the law of total probability, the success probability of the $u$th user can be expressed as
\begin{multline}
s_u(\gamma_0) =\mathbb{E}_{f, \mathbf{x}_u} \Bigg[\sum_{k=1}^{K} \Big[c_{f,k(\mathbf x_u)} \prod_{l=1}^{k-1} \left(1 - c_{f,l(\mathbf{x}_u)}\right)  \Big]   \\
\times \mathbb{P}\left( \gamma_{u,k(\mathbf{x}_u)}(\mathbf{x}_u) > \gamma_0 ~|~ \mathbf{x}_u\right)  \Bigg]  \label{eqn:su}
\end{multline}
where $c_{f,k(\mathbf x_u)}\prod_{l=1}^{k-1} (1 - c_{f,l(\mathbf x_u)}) $ is the probability that the $1$st to the  $(k-1)$th nearest BSs of the $u$th user do not cache the $f$th file and the $k$th nearest BS caches the $f$th file, $\mathbb{P}( \gamma_{u,k(\mathbf x_u)}(\mathbf{x}_u) > \gamma_0 ~ |~ \mathbf{x}_u)$ is the success probability when the user located at $\mathbf x_u$ downloads from the cache of its $k$th nearest BS, and $\mathbb{E}_{f, \mathbf{x}_u}$ denotes the expectation over user request and location.

Similarly, the average achievable rate of the $u$th user can be expressed as
\begin{multline}
\bar R_u = \mathbb{E}_{ f, \mathbf x_u, \mathbf{h}_u}\Bigg[\sum_{k=1}^{K + 1} \Big[c_{f,k(\mathbf x_u)} \prod_{l=1}^{k-1} \left(1 - c_{f,l(\mathbf{x}_u )}\right) \Big]  \\
\times  R_{u,k(\mathbf x_u)}(\mathbf{x}_u) \Bigg]  \label{eqn:Ru}
\end{multline}
where $R_{u,k(\mathbf x_u)}(\mathbf{x}_u) = W_u \log_2(1 + \gamma_{u,k(\mathbf x_u)}(\mathbf{x}_u))$ is the instantaneous data rate of the $u$th user when the user is located at $\mathbf{x}_u$ and downloads from the cache of its $k$th nearest BS, $W_u$ and $\mathbf h_{u} = [h_{u1}, \cdots, h_{uN_b}]$ are the transmission bandwidth for  the $u$th user and channel vector of the user. To unify the expression, we denote the instantaneous data rate when the $u$th user is associated with the local BS to download the file from backhaul as $R_{u,K+1}(\mathbf{x}_u) = \min\{W_u \log_2(1 + \gamma_{u,1(\mathbf x_u)}(\mathbf{x}_u)), C^{\rm bh}_u\}$, where $C^{\rm bh}_u$ is the backhaul bandwidth allocated to the user. Since proactive caching policy is optimized in a much larger time scale (at least in hours) than radio resource allocation (in milliseconds), we do not jointly optimize caching policy and transmission resource allocation.

\begin{proposition}
	The success probability of the $u$th user is
	\begin{multline}
	s_u(\gamma_0) = \sum_{f=1}^{N_f} q_{uf} \sum_{i=1}^{N_b}\sum_{j=1}^{J_i}\frac{a_{ui}|\mathcal{D}_{ij}|}{|\mathcal{D}_i|} \\
	\times \sum_{k=1}^{K} \Big[c_{fk_{ij}} \prod_{l=1}^{k-1} (1 - c_{fl_{ij}})  \Big] {\sf s}_{uk_{ij}}(\gamma_0) \label{eqn:su0}
	\end{multline}
	where $k_{ij}$ denotes the $k$th nearest BS when the user is located in the $j$th region of the $i$th cell $\mathcal{D}_{ij}$, $|\mathcal{D}_{ij}|$ is the area of $\mathcal{D}_{ij}$, $|\mathcal{D}_i|$ is the area of the $i$th cell, $J_i$ is the number of small regions in the $i$th cell, ${\sf s}_{uk_{ij}}(\gamma_0) = \frac{1}{|\mathcal{D}_{ij}|}\iint_{\mathbf{x}_u \in \mathcal{D}_{ij}}$ ${\sf G}_{k_{ij}}(\mathbf{x}_{u}, \gamma_0) {\rm d}x_{u1}{\rm d}x_{u2} $ is the success probability when the user is located within $\mathcal{D}_{ij}$ and downloads from the cache of the $k$th nearest BS, and  ${\sf G}_{k_{ij}}(\mathbf{x}_{u},\gamma_0) = e^{-\gamma_0 ||\mathbf{x}_u - \mathbf{y}_{k_{ij}}||^\alpha\frac{\sigma^2}{P}} \prod_{b\in \Phi_{k_{ij}}}^{ b \neq {k_{ij}}}\big(1 + \gamma_0 \frac{||\mathbf{x}_u - \mathbf{y}_{k_{ij}}||^\alpha}{||\mathbf{x}_u - \mathbf{y}_b||^\alpha}\big)^{-1}$.
\end{proposition}
\begin{IEEEproof}
	See Appendix A.
\end{IEEEproof}

\begin{proposition}
	The average achievable rate of the $u$th user is
	\begin{equation}
	\bar R_u= \sum_{f=1}^{N_f} q_{uf}
	\sum_{i=1}^{N_b}\sum_{j=1}^{J_i}\frac{a_{ui}|\mathcal{D}_{ij}|}{|\mathcal{D}_i|}
	\sum_{k=1}^{K+1} \Big[c_{fk_{ij}} \prod_{l=1}^{k-1} (1 - c_{fl_{ij}})   \Big] {\sf R}_{uk_{ij}}  \label{eqn:Ru0}
	\end{equation}
	where
	\begin{equation}
	{\sf R}_{uk_{ij}}  =
	 \left\{\begin{array}{ll}
	\!\!\!\!\frac{W_u}{|\mathcal{D}_{ij}|} \iint_{\mathbf{x}_u \in \mathcal{D}_{ij}} \big[\delta_{uk_{ij}k_{ij}} {\sf F}_{k_{ij}}(\mathbf{x}_u ) + 
	   \sum_{b\in \Phi_{k_{ij}}, b\neq k_{ij}}  \\  
	   \quad\quad(\delta_{ubk_{ij}}-\delta_{ub\bar k_{ij}}) {\sf F}_b(\mathbf{x}_u)\big]{\rm d}x_{u1}{\rm d}x_{u2},~\text{if}~k\leq K  \\
	\!\!\!\!\frac{1}{|\mathcal{D}_{ij}|} \iint_{\mathbf{x}_u \in \mathcal{D}_{ij}}  
	\left[\int_{0}^{C_u^{\rm bh}} {\sf G}_{1_{ij}}(\mathbf{x}_{u}, 2^{\frac{t}{W_u} - 1}) {\rm d}t \right]
	\\\quad\quad{\rm d}x_{u1}{\rm d}x_{u2} ,~\text{if}~k = K + 1
	\end{array}
	\right.
	\end{equation}
	is the average achievable rate when the user is located within $\mathcal{D}_{ij}$ and downloads from the cache of the $k$th nearest BS (or from the backhaul if $k= K + 1$), $\delta_{ubk_{ij}} =\prod_{b'\in\Phi_{k_{ij}}}^{ b'\neq b} \frac{r_{ub'}^{\alpha}}{r_{ub'}^{\alpha} - r_{ub}^{\alpha}}$,  $\delta_{ub\bar k_{ij}} = \prod_{b'\in\Phi_{k_{ij}}}^{b'\neq b,k_{ij}} \frac{r_{ub'}^{\alpha}}{r_{ub'}^{\alpha} - r_{ub}^{\alpha}} $,
$
{\sf F}_b(\mathbf{x}_u) =-\frac{\exp({\frac{\sigma^2}{P} ||\mathbf{x}_u - \mathbf{y}_b||^\alpha})}{\ln 2} {\rm Ei}(- \tfrac{\sigma^2}{P}||\mathbf{x}_u - \mathbf{y}_b||^\alpha) + \log_2 \frac{\sigma^2}{P}
$, and ${\rm Ei}(x)= -\int_{-x}^{\infty}\frac{e^{-t}}{t}dt$ is the exponential integral.
\end{proposition}
\begin{IEEEproof}
	See Appendix B.
\end{IEEEproof}

The success probability and average achievable rate of each user in the propositions are in closed-form with respect to the caching probabilities, user location probabilities and preferences, which enables us to optimize caching policy and analyze the impact of spatial locality and preference heterogeneity. Although ${\sf s}_{uk_{ij}}(\gamma_0)$ and ${\sf R}_{uk_{ij}}$ in \eqref{eqn:su0} and \eqref{eqn:Ru0} contain numerical integrals, their values only depend on network configurations such as BS locations, SINR threshold, transmit power, noise power, and pathloss exponent. When we optimize the caching policy for a given network, these two terms can be treated as constant after being computed.

\begin{figure}[!htb]
	\centering	
	\subfigure[Relation between ${\sf s}_{uk}(\gamma_0)$ and $k$]{
		\label{fig:suc_vs_gamma} 
		\includegraphics[width=0.23\textwidth]{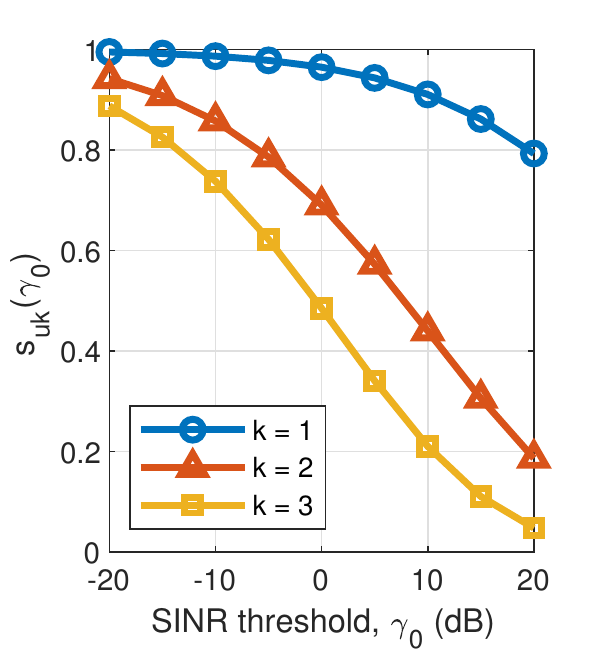}}
	\subfigure[Relation between ${\sf R}_{uk}$ and $k$]{
		\label{fig:R_vs_k} 
		\includegraphics[width=0.23\textwidth]{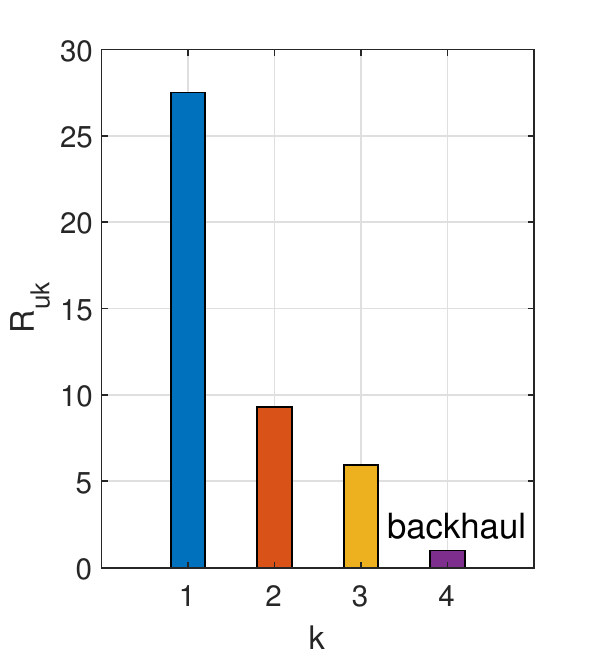}}
	\caption{Numerical results of the two terms ${\sf s}_{uk}(\gamma_0)$ and ${\sf R}_{uk}$, both quickly reduce with $k$. }
	\label{fig:numerical}
\end{figure}

When considering the hexagonal cell model shown in Fig.~\ref{fig:layout}, the two terms do not depend on $i$ and $j$ due to the symmetry of the BS topology. Then, we only need to compute ${\sf s}_{uk_{ij}}(\gamma_0)$ and ${\sf R}_{uk_{ij}}$ when $i=j=1$, i.e., when the user is located in the shaded area $\mathcal{D}_{11}$ of Fig.~\ref{fig:layout}, without loss of generality. Since ${\sf s}_{uk_{ij}}(\gamma_0) = {\sf s}_{uk_{11}}(\gamma_0)$ and ${\sf R}_{uk_{ij}} = {\sf R}_{uk_{11}}$ for $\forall i,j$, we can use notations ${\sf s}_{uk}(\gamma_0)$ and ${\sf R}_{uk}$ to replace ${\sf s}_{uk_{ij}}(\gamma_0)$ and ${\sf R}_{uk_{ij}}$. Besides, with the hexagonal model, the integral domain $\iint_{\mathbf{x}_u \in \mathcal{D}_{11}}$ for computing ${\sf s}_{uk}(\gamma_0)$ and ${\sf R}_{uk}$ can be explicitly expressed as $\int_{0}^{D}\!\!\int_{0}^{\frac{x_{u2}}{\sqrt{3}}}$, and the area of $\mathcal{D}_{ij}$  can be obtained as $|\mathcal{D}_{ij}| = \frac{D^2}{2\sqrt{3}}$, where $D$ denotes the cell radius. Then, the integrals can be numerically computed, e.g., by the built-in function {\tt integral2} in MATLAB.\footnote{For arbitrary given BS topology, the small region $\mathcal{D}_{ij}$ becomes irregular so that the integral domain is hard to be expressed explicitly. Nevertheless, the integrals together with areas $|\mathcal D_{ij}|$ and $|\mathcal{D}_i|$ can be computed by Monte Carlo method~\cite{caflisch1998monte}.} In Fig.~\ref{fig:numerical}, we show the numerical results of the two terms under typical network configurations given in Section~V, which suggest that $K$ can be set as a small number, i.e., the size of the neighboring BS set is small.

\subsection{Caching Policy Optimization}
 We can see from Propositions 1 and 2 that $s_u(\gamma_0)$ and $\bar R_u$ have same function structure with respect to $\{c_{fb}\}$. To unify the optimization framework, we introduce a \emph{user utility} function as
\begin{equation}
T_u \triangleq  \sum_{f=1}^{N_f} q_{uf} \sum_{i=1}^{N_b}\sum_{j=1}^{J_i}\frac{a_{ui} |\mathcal{D}_{ij}|}{|\mathcal D_{i}|}\sum_{k=1}^{K + 1} \Big[(c_{fk_{ij}} \prod_{l=1}^{k-1} (1 - c_{fl_{ij}})   \Big] {\sf T}_{uk_{ij}} \label{eqn:Tuij}
\end{equation}
 where ${\sf T}_{uk_{ij}}$ denotes the utility of the $u$th user when the user downloads from the cache of its $k$th nearest BS (or from the backhaul if $k = K+1$) when located in $\mathcal{D}_{ij}$. ${\sf T}_{uk_{ij}}$ can either be ${\sf s}_{uk_{ij}}(\gamma_0)$ or  ${\sf R}_{uk_{ij}}$, and then $T_{u}$ represents either the success probability $s_u(\gamma_0)$ or the average achievable rate $\bar R_u$ of the $u$th user accordingly. For the hexagonal model, \eqref{eqn:Tuij} degenerates into
\begin{equation}
T_u =  \sum_{f=1}^{N_f} q_{uf}\sum_{i=1}^{N_b}\sum_{j=1}^{12}\frac{a_{ui}}{12}\sum_{k=1}^{K + 1} \Big[(c_{fk_{ij}} \prod_{l=1}^{k-1} (1 - c_{fl_{ij}})   \Big] {\sf T}_{uk} \label{eqn:Tu}
\end{equation}
Since \eqref{eqn:Tu} and \eqref{eqn:Tuij} have the same function structure, we only consider the hexagonal model in the following for notational simplicity, but the results are applicable to arbitrary BS distribution.

From the network perspective, we employ  \emph{network utility} as a metric to reflect the average user experience of all users in the network as
\begin{multline}
T =\mathbb{E}_u [T_u] =\sum_{u=1}^{N_u} v_u T_u = \sum_{u=1}^{N_u}v_u\sum_{f=1}^{N_f}   q_{uf}\sum_{i=1}^{N_b}\sum_{j=1}^{12}\frac{a_{ui}}{12} \\
\times \sum_{k=1}^{K + 1} \Big[c_{fk_{ij}} \prod_{l=1}^{k-1} (1 - c_{fl_{ij}})   \Big] {\sf T}_{uk}  \label{eqn:R}
\end{multline}
where $T$ represents the \emph{network success probability} or the \emph{network average rate}, which is the success probability or the average achievable rate averaged over all the content requests in the considered region. If we set $\gamma_0 = 0$, then the network success probability degenerates into the \emph{cache-hit probability}, another often-used metric in the literature of caching~\cite{czy}.

From the user fairness perspective, we consider max-min fairness~\cite{maxmin} to reflect the worst user experience, which lets every user benefit from caching in its own experience. This is achieved by maximizing the minimal utility among all the users in the $N_b$ cells.

Both network average utility and minimal user utility can reflect user experience, but from different perspectives. To provide the flexibility for a cache manager in balancing the two metrics, we formulate the following optimization problem that maximizes their weighted sum
\begin{subequations}
	\begin{align}
	{\sf P}_0:	~~\max_{\{c_{fb}\}} ~& (1-\eta) T + \eta\min_{u=1,\cdots,N_u}\left\{ T_u \right\} \label{eqn:obj}\\
	s.t. ~&\sum_{f=1}^{N_f} c_{fb} \leq N_c,~ \forall b \label{eqn:size}\\
	& 0\leq c_{fb} \leq 1,~ \forall f, b  \label{eqn:pro}
	\end{align}
\end{subequations}
where \eqref{eqn:size} is equivalent to the cache size constraint for probabilistic caching policy~\cite{Blaszczyszyn2015optimal}, and \eqref{eqn:pro} is the probability constraint. The value of weight $\eta$ depends on how the cache manager (e.g., the MNO) trades off between network average performance and user fairness. By setting $\eta$ as $0$ or $1$, we can obtain a problem of maximizing the network utility (referred to as Problem ${\sf P}_1$) or a problem maximizing the minimal utility among users (referred to as Problem ${\sf P}_2$).

Based on variable replacement $z_{fb} \triangleq 1 - c_{fb}$, the expressions of the two metrics can be rewritten as
$
T_u ={\sf T}_{u1} -\frac{1}{12} \sum_{f=1}^{N_f}\sum_{i=1}^{N_b}\sum_{j=1}^{12}a_{ui}q_{uf} \sum_{k=1}^{K} (\mathsf{T}_{uk} - \mathsf{T}_{u,K+1})\prod_{l=1}^{k} z_{fl_{ij}}
$
and
$
\min_{u=1,\cdots,N_u}\{T_u\} = {\sf T}_{u1} - \frac{1}{12}\max_{u=1,\cdots,N_u} \{\sum_{f=1}^{N_f}\sum_{i=1}^{N_b}\sum_{j=1}^{12}a_{ui}q_{uf} \sum_{k=1}^{K} (\mathsf{T}_{uk} - \mathsf{T}_{u,K+1})\prod_{l=1}^{k} z_{fl_{ij}} \}
$,
respectively. By further introducing an auxiliary variable $t$ and extra constraints,  we can convert the original problem ${\sf P}_0$ equivalently into
\begin{subequations}
	\begin{align}
{\sf SP}:	\min_{\{z_{fb}\}, t} & (1-\eta) \sum_{u=1}^{N_u}\sum_{f=1}^{N_f}\sum_{i=1}^{N_b}\sum_{j=1}^{12} v_ua_{ui}q_{uf} \nonumber \\
& \times \sum_{k=1}^{K} (\mathsf{T}_{uk} - \mathsf{T}_{u,K+1})\prod_{l=1}^{k} z_{fl_{ij}} + \eta t  \label{eqn:obj1}\\
	s.t. ~& \sum_{f=1}^{N_f}\sum_{i=1}^{N_b}\sum_{j=1}^{12}a_{ui}q_{uf} \sum_{k=1}^{K} (\mathsf{T}_{uk} - \mathsf{T}_{u,K+1}) \nonumber \\
 &	\times	\prod_{l=1}^{k} z_{fl_{ij}}  \leq t, ~\forall u
	\label{eqn:conmax} \\
	&  N_f - \sum_{f=1}^{N_f} z_{fb} \leq N_c,~ \forall b \label{eqn:consize} \\
	& 0\leq z_{fb} \leq 1,~ \forall f, b \label{eqn:prob}
	\end{align}
\end{subequations}
Since the utility of a user when downloading from the cache at the $k$th nearest BS is larger than that at the $(k+1)$th nearest BS due to shorter BS-to-user distance, we have ${\sf T}_{uk} \geq {\sf T}_{u,K+1}$. Hence, the objective function is a posynomial function\footnote{A posynomial function is with the form $f(\mathbf{x}) = \sum_{k=1}^{K} \psi_k x_1^{\phi_{1k}} x_2^{\phi_{2k}} \cdots x_n^{\phi_{nk}} $, where $\mathbf{x} =[x_1, \cdots, x_n] \in \mathbb{R}^{n}$, $\psi_k \geq 0$ and $\phi_{ik} \in \mathbb{R}$. If there exists $\psi_k < 0$, then $f(\mathbf{x})$ is a signomial function~\cite{boyd2004convex}.} with respect to $\{z_{fb}\}$ and $t$. Constraint \eqref{eqn:conmax} can be rewritten as $\sum_{f=1}^{N_f}\sum_{i=1}^{N_b}\sum_{j=1}^{12}a_{ui}q_{uf} \sum_{k=1}^{K} (\mathsf{T}_{uk} - \mathsf{T}_{u,K+1})t^{-1}\prod_{l=1}^{k} z_{fl_{ij}}  \leq 1$, where the left hand side is also a posynoimal function. However, the left hand side of constraint \eqref{eqn:consize} is a signomial function due to the negative sign before the term $\sum_{f=1}^{N_f} z_{fb}$. Therefore, ${\sf SP}$ is a signomial programming problem, which is truly nonconvex~\cite{Chiang}.

To solve the problem, we first replace \eqref{eqn:consize} by a posynomial function based on the arithmetic-geometric inequality. Given a set of non-negative weights $\{\varepsilon_{fb}^{(n)}\}$ with $\sum_{f = 1}^{N_f} \varepsilon_{fb}^{(n)} = 1$, the arithmetic-geometric inequality gives
$\sum_{f = 1}^{N_f}z_{fb} \geq \prod_{f = 1}^{N_f} \big({z_{fb}}/{\varepsilon_{fb}^{(n)}}\big)^{\varepsilon_{fb}^{(n)}}
$,
where the equality holds if and only if $\varepsilon_{fb}^{(n)} = {z_{fb}}/{\sum_{f=1}^{N_f} z_{fb}}$~\cite{Chiang}. Then, any set of variables $\{z_{fb}\}$ satisfying a more strict constraint
$
N_f - \prod_{f = 1}^{N_f} \big({z_{fb}}/{\varepsilon_{fb}^{(n)}}\big)^{\varepsilon_{fb}^{(n)}} \leq N_c
$
will also satisfy \eqref{eqn:consize}. By replacing \eqref{eqn:consize} with such a more strict constraint and after some manipulations, we can obtain a condensed problem as
\begin{subequations}
	\begin{align}
	{\sf GP}^{(n)}:	\min_{\{z_{fb}\}, t} & (1-\eta) \sum_{u=1}^{N_u}\sum_{i=1}^{N_b}\sum_{j=1}^{12} \sum_{f=1}^{N_f}v_ua_{ui}q_{uf} \nonumber \\
	&\times \sum_{k=1}^{K} (\mathsf{T}_{uk} - \mathsf{T}_{u,K+1})\prod_{l=1}^{k} z_{fl_{ij}} + \eta t  \\
	s.t. ~&(N_f - N_c) \prod_{f = 1}^{N_f}\! \left({z_{fb}}/{\varepsilon_{fb}^{(n)}}\right)^{-\varepsilon_{fb}^{(n)}} \!\!\leq 1,~ \forall b \label{eqn:contight} \\
	& \eqref{eqn:conmax}, \eqref{eqn:prob}\nonumber
	\end{align}
\end{subequations}
where any feasible solution of ${\sf GP}^{(n)}$ is a feasible solution of ${\sf SP}$. ${\sf GP}^{(n)}$ is a geometric programming problem since the objective function and all the constraints are posynomial functions. By taking variable replacements $\tilde z_{fb} = \ln z_{fb}$, $\tilde t = \ln t$  and logarithmic transformation on the objective function and constraints, we can convert ${\sf GP}^{(n)}$ into a convex problem~\cite{boyd2004convex} whose global optimal solution  can be found by standard convex optimization method, say interior-point method  whose computational complexity is in the order of $\mathcal{O}((N_uN_bN_f)^{3.5})$~\cite{nemirovski2004interior}.

Since ${\sf GP}^{(n)}$ can serve as an accurate approximation of ${\sf SP}$ when $\varepsilon_{fb}^{(n)} \approx {z_{fb}}/{\sum_{f=1}^{N_f} z_{fb}}$, to improve the accuracy, the value of $\varepsilon_{fb}^{(n)}$ should be updated iteratively by solving a series of problems ${\sf GP}^{(1)}, {\sf GP}^{(2)},\cdots$. Let $\{ z_{fb}^{(n)} \}$ denote the optimal solution of ${\sf GP}^{(n)}$. By successively updating $\varepsilon_{fb}^{(n)} = z_{fb}^{(n-1)}/\sum_{f=1}^{N_f} z_{fb}^{(n-1)}$,  we can obtain $\lim\limits_{n\to \infty}\varepsilon_{fb}^{(n)} = \lim\limits_{n\to \infty} z_{fb}^{(n)}/\sum_{f=1}^{N_f} z_{fb}^{(n)}$, which suggests that the approximation is accurate (and hence ${\sf GP}^{(n)}$ is equivalent to the original ${\sf SP}$) within the neighborhood of  $\lim\limits_{n\to \infty} z_{fb}^{(n)}$. In fact, point $\lim\limits_{n\to \infty} z_{fb}^{(n)}$ obtained by such successive approximation method is proved to satisfy the Karush-Kuhn-Tucher (K.K.T) condition of the original ${\sf SP}$ and is reported to converge to the global optimal solution of ${\sf SP}$ in most experiments~\cite{Chiang}.\footnote{Our simulations show that Algorithm 1 can converge to, at least, a local optimal solution. Since ${\sf SP}$ is an intractable NP-hard problem~\cite{Chiang}, it is hard to verify whether or not the global optimal solution is found in the large-scale problem as in our case.} Finally, considering the equivalence between ${\sf P}_0$ and ${\sf SP}$, we can obtain $c_{fb}^* =  1- \lim\limits_{n\to \infty}z_{fb}^{(n)}$ as an optimal solution of the original problem ${\sf P}_0$.  The whole procedure of solving ${\sf P_0}$ is given in Algorithm~1. The computational complexity of Algorithm~1 is  $\mathcal{O}(L(N_uN_bN_f)^{3.5})$, where $L$ is number of iterations for $z_{fb}^{(n)}$ to converge that is not large as shown by simulations in Section V.

\begin{algorithm}
	\caption{ Solving Problem ${\sf P}_0$} \label{al:1}
		\begin{algorithmic}[1]
			\REQUIRE An initial feasible caching solution $\{c_{fb}^{(0)}\}$, e.g., ${c}_{fb}^{(0)} = {N_c}/{N_f}$, and error tolerance $\epsilon$.
			\ENSURE Optimal caching policy $ \{c_{fb}^*\}$
			\STATE $ z_{fb}^{(0)} =1 - c_{fb}^{(0)} $,  initialize $ z_{fb}^{(1)} = {\sf inf}$ and  $n = 1$
			\WHILE{ $|| z_{fb}^{(n)} - z_{fb}^{(n - 1)} || > \epsilon$  }
			\STATE Update $\varepsilon_{fb}^{(n)} = z_{fb}^{(n-1)}/\sum_{f=1}^{N_f} z_{fb}^{(n-1)}$
			\STATE Obtain the optimal solution of ${\sf GP}^{(n)}$, i.e, $\{z_{fb}^{(n)}\}$,  by solving the converted convex problem using interior-point method.
			\STATE $n\leftarrow n + 1$
			\ENDWHILE
			\STATE $ c_{fb}^{*} =1 - z_{fb}^{(n)}$
		\end{algorithmic}
\end{algorithm}


\subsection{Analysis for Special Cases}
To reveal how spatial locality, user preference heterogeneity, and different performance metrics affect the caching policy, we analyze the optimal solutions of Problem ${\sf P}_1$ and Problem ${\sf P}_2$, respectively, which are referred to as \emph{Policy~1} and \emph{Policy~2} in the following.

\begin{corollary}
	Policy~1 satisfies $c_{fb}^* \in \{0, 1\}$.
\end{corollary}
\begin{IEEEproof}
Suppose that there exists $c^*_{f'b'}$ satisfying $0<c^*_{f'b'}<1$, e.g., $0<c^*_{11}<1$. If we fix the value of $\{c_{fb}^*\}_{f = 1, \cdots, N_f, b = 2,\cdots, N_b}$ and optimize $\{c_{f1}\}_{f=1,\cdots, N_f}$, then the objective function $T$ can be rewritten as  $\sum_{f=1}^{N_f} \zeta_f c_{f1} + \zeta_0$ by further considering \eqref{eqn:R}, where $\{\zeta_f\}_{f = 1, \cdots,N_f}$ are constants that do not depend on $\{c_{f1}\}_{f=1,\cdots, N_f}$. Then, the problem maximizing the network utility can be reformulated as $\max_{\{c_{f1}\}} \sum_{f=1}^{N_f} \zeta_f c_{f1}  +  \zeta_0$ subject to  $\sum_{f=1}^{N_f} c_{f1}\! \leq\! N_c$ and $ 0\!\leq c_{f1}\! \leq 1,~ \forall f$,~which~is a linear programming problem. We can easily see that the optimal solution is $c^*_{fb} = 1$ if $n(f) \leq N_c$ and $c^*_{fb} = 0$ if $n(f) > N_c$, where $n(f)$ denotes the ranking of the value $\zeta_f$ in $\{\zeta_1, \cdots, \zeta_{N_f}\}$ in descending order. Therefore, $c_{11}^* \in \{0, 1\}$, which contradicts with $0<c^*_{11}<1$.
\end{IEEEproof}

Corollary 1 means that when only maximizing the network utility (e.g., network average rate or equivalently average sum rate, and cache-hit probability), probabilistic caching policy degenerates into deterministic caching policy as designed in~\cite{femtocachingTIT,liujuan,CBQ}. The result is due to the fixed BS topology. When considering user fairness, probabilistic policy should be employed.

To analyze when exploiting user preference is beneficial, we denote the network utility based on global content popularity as $\tilde T$. It is obtained by replacing $q_{uf}$ in \eqref{eqn:R} with $p_f$ as $\tilde T =\sum_{u=1}^{N_u}v_u \sum_{f=1}^{N_f} p_{f} T_{uf}$, where $T_{uf} \triangleq \sum_{i=1}^{N_b} \sum_{j=1}^{12}\frac{a_{ui}}{12}\sum_{k=1}^{K + 1} \big[(c_{fk_{ij}} $ $\prod_{l=1}^{k-1} (1 - c_{fl_{ij}})   \big] {\sf T}_{uk}$ is the utility of the $u$th user when downloading the $f$th file. Considering \eqref{eqn:relation}, we have
$\tilde T = \mathbb{E}_u [\sum_{f=1}^{N_f} p_f T_{uf}] =\sum_{f=1}^{N_f}  p_f \mathbb{E}_u [ T_{uf}] = \sum_{f=1}^{N_f} \mathbb{E}_u [q_{uf}] \mathbb{E}_u [T_{uf}] $. From \eqref{eqn:R}, the network utility based on user preference can be expressed as $T = \mathbb{E}_u [\sum_{f=1}^{N_f} q_{uf} T_{uf}] = \sum_{f=1}^{N_f}\mathbb{E}_u [ q_{uf} T_{uf}]$. Then, we can obtain the following corollary.

\begin{corollary}
	The network utility only depends on global content popularity (i.e., $T = \tilde T$) if and only if $q_{uf}$ and $T_{uf}$ are uncorrelated with respect to $u$ (i.e., $\mathbb{E}_u [ q_{uf} T_{uf}] = \mathbb{E}_u [q_{uf}] \mathbb{E}_u [T_{uf}]$).
\end{corollary}

When users preferences are homogeneous (i.e., $q_{uf} = p_f$), $q_{uf}$  does not depend on $u$ and hence is uncorrelated with $T_{uf}$. Or, when users are without spatial locality and with identical transmission resource (i.e., $a_{ui} = 1/N_b$ and  ${\sf T}_{1k} = \cdots = {\sf T}_{N_uk}$), we have $T_{uf}= T_{1f} = \cdots = T_{N_uf}$, which does not depend on $u$ and hence is uncorrelated with $q_{uf}$. In both cases, using the knowledge of global content popularity is sufficient for optimizing caching policy to achieve the maximal network utility, i.e., exploiting user preference does not yield better caching policy.

To provide insight for the impact of objective and user behavior on caching policy, in the following corollaries we consider a case where ${\sf T}_{1k} = \cdots = {\sf T}_{N_uk} \triangleq {\sf T}_k$, i.e., the utility
of every user when downloading from the cache of its $k$th nearest BS is identical. This implies identical transmission resource for each user as shown from the expression of ${\sf R}_{uk}$. In this case, $T$ in \eqref{eqn:R} degenerates into a function of $p_{fi}$, i.e., using the knowledge of local content popularity is sufficient for optimizing caching policy to achieve the maximal value of network utility.

\begin{corollary}
When the user utility of downloading from local BS's cache far exceeds that from the second nearest BS's cache, Policy~1 is $c_{fb}^* = 1$ if $n(f,b) \leq N_c$, and $c_{fb}^* = 0$ if $n(f,b) > N_c$, where $n(f,b)$ denotes the ranking of the value $p_{fb}$ in $\{p_{1b},\cdots,p_{N_fb}\}$ in descending order.
\end{corollary}
\begin{IEEEproof}
When ${\sf T}_1 \gg {\sf T}_2$, $T$ in \eqref{eqn:R} degenerates into
$
\lim_{{\sf T}_2/{\sf T}_1 \to 0}T = \frac{{\sf T}_1}{12} \sum_{u,f,j,i} a_{ui}v_uq_{uf}\big(  c_{fi} + \sum_{k=2}^{K + 1} \big[c_{fk_{ij}} \prod_{l=1}^{k-1} (1 - c_{fl_{ij}})   \big] \frac{{\sf T}_{k}}{{\sf T}_{1}} \big) = {\sf T}_1\sum_{u,f,i} a_{ui}v_uq_{uf}   c_{fi}
$,
from which we can see that maximizing $T$ subject to \eqref{eqn:size} and \eqref{eqn:pro} is equivalent to maximizing  $\sum_{u,f} a_{ui}v_uq_{uf} c_{fi}$ for $\forall j$ subject to $\sum_{f=1}^{N_f}c_{fi}\leq N_c$ and $0\leq c_{fi}\leq1$. Considering \eqref{eqn:local}, we have $\sum_{u,f} a_{ui}v_uq_{uf} c_{fi} = (\sum_{u = 1}^{N_u}a_{ui}v_u)\sum_{f=1}^{N_f} p_{fi} c_{fi} $. Then, Policy~1  is to let the $i$th BS to cache the $N_c$ files with the largest values of $p_{fi}$.
\end{IEEEproof}

In typical wireless networks, when regarding average achievable rate as the utility (i.e., $\mathsf T_k = \mathsf R_{uk}$), ${\sf T}_1$ is much larger than ${\sf T}_2$ as shown in Fig.~\ref{fig:R_vs_k} due to shorter BS-to-user distance. Then, Corollary 3 suggests that Policy~1 tends to cache the most locally popular files at each BS. When choosing success probability as the utility (i.e., $\mathsf T_k = \mathsf s_{uk}(\gamma_0)$), ${\sf T}_1$ and ${\sf T}_2$ can be close if the SINR threshold $\gamma_0$ is not high (e.g., $-5$ dB) as shown in Fig.~\ref{fig:suc_vs_gamma}. Then, Policy~1 tends to cache the locally popular files in the neighboring BS sets more distributively, which can increase cache-hit probability by file diversity (i.e., content diversity~\cite{chenzhen}).




\begin{corollary}
	If the spatial location probabilities, activity level and preferences are identical for all users, Policy~1 will be the same as Policy~2.
\end{corollary}

\begin{IEEEproof}
If $\mathbf{a}_1 = \cdots =\mathbf{a}_{N_u}$, $v_u=1/N_u$ and $q_{uf}=p_f$, $T$ in \eqref{eqn:R} will degenerate into $T = T_u, \forall u$ and hence $T = \min_{u=1,\cdots, N_u}\{T_u\}$. Then, problems ${\sf P}_1$ and ${\sf P}_2$ are equivalent.
\end{IEEEproof}
Corollary 4 suggests that if one assumes that each user has no difference in spatial locality, activity level and preference (and with  identical transmission resources) (i.e., there is no difference among users statistically), maximizing the network utility is equivalent to maximizing the minimal utility among users. In practice, these assumptions are hardly true. This implies a tradeoff between maximizing network utility and maximizing user fairness.

\subsection{Numerical Examples}
To help understand the impact of heterogeneous user preference given that users are with spatial locality on the behavior of Policy~1 and Policy~2, we consider two toy examples respectively with one BS and two BSs, and consider average rate as the utility. There are two user equipments (UEs) and three files. The global popularity is  $\mathbf p = [0.46, 0.30, 0.24]$ and the activity level distribution is $\mathbf v = [0.6, 0.4]$. For each example, we compare two cases with homogeneous user preference ($\mathbf q_1 = \mathbf q_2 = \mathbf{p} = [0.46, 0.30, 0.24]$) and heterogeneous user preference ($\mathbf q_1 = [0.75, 0.25, 0], \mathbf q_2 = [0.02, 0.38, 0.60]$), respectively, where the user preference satisfies the relation $ v_1 \mathbf q_1 + v_2 \mathbf q_2 = \mathbf p$ given by \eqref{eqn:relation}. The cache size is $N_c = 1$. The caching policy of the $b$th BS is denoted as $\mathbf{c}_b = [c_{1b},c_{2b},c_{3b}]$.

{\it 1) Single-cell:}
In this case, both users are located in the same cell and the user location probability matrix becomes $\mathbf{A} = \left[1, 1\right]^T$.

\begin{figure}[!htb]
	\centering
	\subfigure[Homogeneous, Policy~1.]{
		\label{fig:11hom} 
		\includegraphics[width=0.23\textwidth]{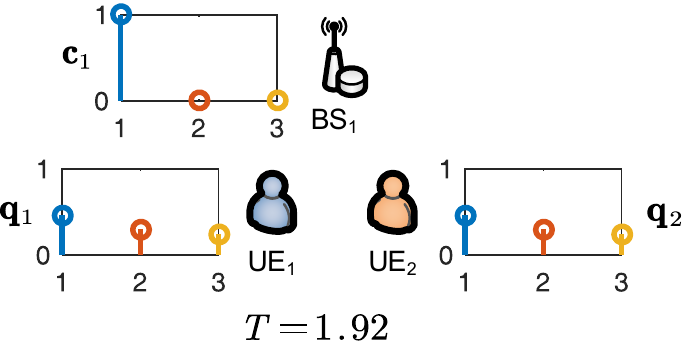}}
	\subfigure[Homogeneous, Policy~2.]{
		\label{fig:12hom} 
		\includegraphics[width=0.23\textwidth]{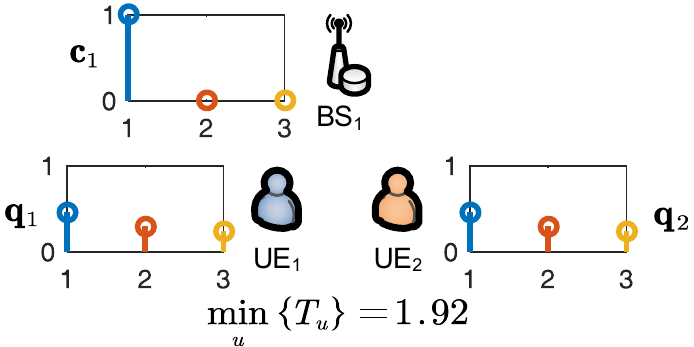}}
	\subfigure[Heterogeneous, Policy~1.]{
		\label{fig:11} 
		\includegraphics[width=0.23\textwidth]{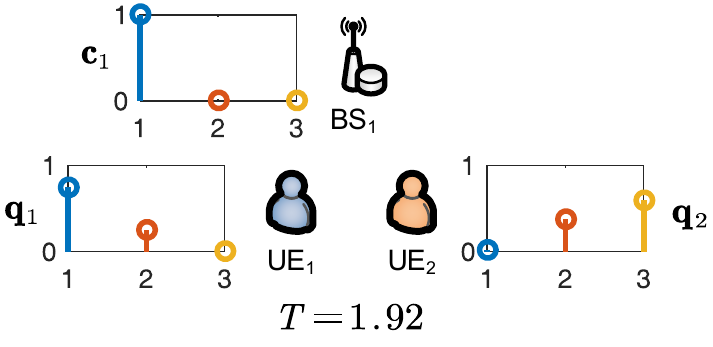}}
	\subfigure[Heterogeneous, Policy~2.]{
		\label{fig:12} 
		\includegraphics[width=0.23\textwidth]{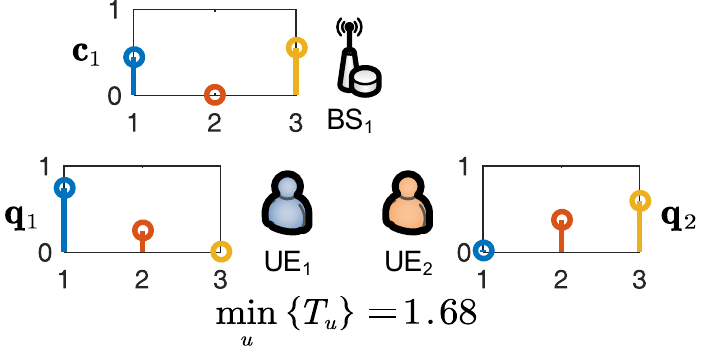}}
	\caption{Two UEs are located in one cell. The user utilities when downloading from the cache and backhaul are 3 Mbps and 1 Mbps, respectively. }
	\label{fig:example1} 
\end{figure}

The optimization results are given in Fig. \ref{fig:example1}. We can see  that when user preferences are homogeneous, both policies let BS$_1$ cache the most preferable file of UE$_1$ and UE$_2$, i.e., file~1.

When user preferences become heterogeneous, the most preferred file of UE$_1$ and UE$_2$ are file~1 and file~3, respectively. Therefore, UE$_1$ prefers BS$_1$ to cache file~1 so that its utility can be maximized while UE$_2$ prefers BS$_1$ to cache file~3, i.e., the \emph{caching interests} of both UEs conflict with each other. Since UE$_1$ is more active than UE$_2$, which results in higher local content popularity for file~1, Policy~1 lets BS$_1$ cache file~1, which agrees with Corollary 3.  This, however, sacrifices the utility of UE$_2$ and makes UE$_2$ achieving the minimal utility. Therefore, to ensure max-min fairness, Policy~2  allocates non-zero probability to BS$_1$ to cache the most preferable file of UE$_2$, i.e., file~3. Compared with the case of homogeneous user preference, the minimal utility decreases due to the conflict of users' caching interests.

{\it 2) Two-cell:}
In this case, each cell has one user and $\mathbf{A} =\left[\begin{smallmatrix}1&0\\0&1\end{smallmatrix}\right]$.

\begin{figure}[!htb]
	\centering
	\subfigure[Homogeneous, Policy~1.]{
		\label{fig:21hom} 
		\includegraphics[width=0.23\textwidth]{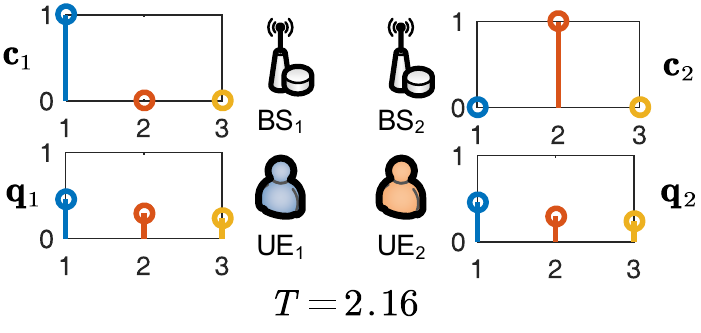}}
	\subfigure[Homogeneous, Policy~2.]{
		\label{fig:22hom} 
		\includegraphics[width=0.23\textwidth]{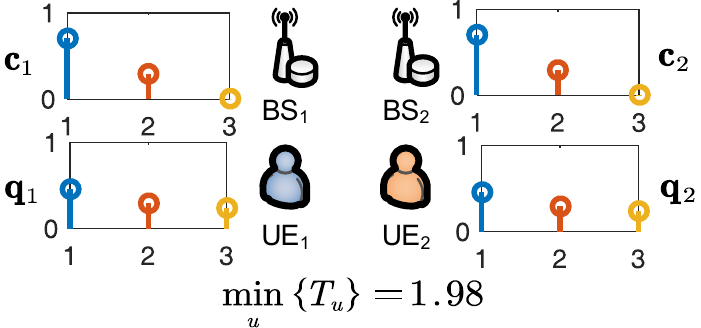}}
	\subfigure[Heterogeneous, Policy~1.]{
		\label{fig:21} 
		\includegraphics[width=0.23\textwidth]{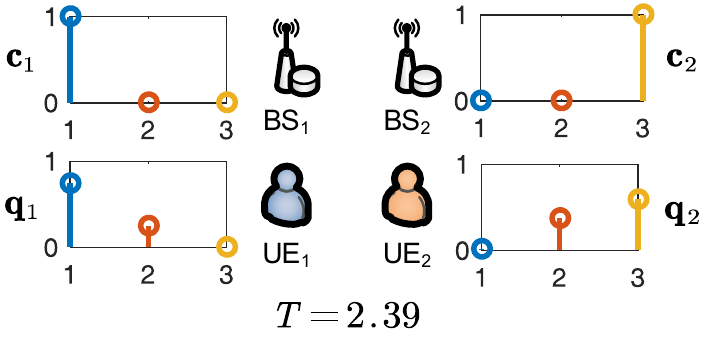}}
	\subfigure[Heterogeneous, Policy~2.]{
		\label{fig:22} 
		\includegraphics[width=0.23\textwidth]{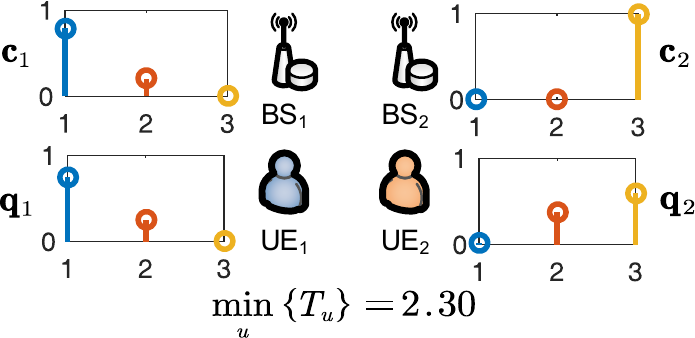}}
	\caption{Two UEs are located in two cells. The utilities of both UEs when downloading from the caches at the local BS, neighboring BS, and backhaul are 3, 2, and 1 Mbps, respectively. }
	\label{fig:example2} 
\end{figure}

The optimization results are given in Fig.~\ref{fig:example2}. In this case, it is interesting to see that when user preferences are homogeneous, the caching interests of both users are exactly the opposite, i.e., UE$_1$ prefers its local BS (i.e., BS$_1$) to cache its most preferable file (i.e., file~1) and its neighboring BS (i.e., BS$_2$) to cache its second preferable file (i.e., file~2), while UE$_2$ prefers BS$_2$ to cache file~1 and BS$_1$ to cache file~2. Since UE$_1$ has higher activity level, Policy~1 lets BSs cache  files according to UE$_1$'s cache interest, which scarifies the utility of UE$_2$. Policy~2 lets each BS cache the most preferable file (i.e., file~1) of its local user with a higher probability and cache the second preferable file (i.e., file~2) of its local user with a lower probability.

When user preferences are heterogeneous, UE$_1$ prefers BS$_1$ to cache file~1 and BS$_2$ to cache file~2, while UE$_2$ prefers BS$_2$ to cache file~3 and BS$_1$ to cache file~2. As a result, Policy~1 lets each BS cache the most preferable file of its local user, i.e., BS$_1$ caches file~1 and BS$_2$ caches file~3. Since UE$_2$ has the minimal utility, Policy~2 is more prone to let BSs cache the files preferred by UE$_2$, i.e., allocating non-zero probability to BS$_1$ to cache the second preferable file of UE$_2$ (i.e., file~2). Compared with the case of homogeneous user preference, both the network utility and minimal utility increase, which can be explained as follows.

\begin{enumerate}
	\item When user preferences are heterogeneous, the most favorable files of the users located in different cells differ. File diversity can be achieved  by simply letting each BS satisfy the caching interest of its local user, which increases the cache-hit probability since files are cached less redundantly in the neighboring BS set.
	\item With given content popularity, the preferences of both users are more skewed when the preferences are less similar.
	Skewed user preference means less uncertain user behavior in requesting files, which amplifies the gain of file diversity.
\end{enumerate}

The two examples show that the heterogeneity of user preferences is a double-edged sword. On one hand, the caching interests of users located in the same cell conflict with each other when user preferences become heterogeneous, which degrades user fairness. On the other hand, the caching interests of users located in different cells are less conflicting when user preferences are heterogeneous, so that file diversity can be achieved with less scarifice of users' caching interests. Besides, the uncertainty of user demands reduces for a given content popularity, which is beneficial for both network utility and user fairness.

\section{Real Datasets Analysis and User Preference Synthesis}
In this section, we first examine the common assumptions on user activity level and user preference, and analyze preference similarity based on two real datasets.
To reveal the entangled impact of user behavior in different aspects on caching gain, we then provide an algorithm to synthesize user preferences with given content popularity, user activity level and controlled average similarity. Finally, we validate the algorithm by the datasets.

\subsection{User Behavior Analysis with Real Datasets} \label{subsec:a}

We use \emph{Million Songs Dataset (MSD)}\cite{MSD} and \emph{Lastfm-1K Dataset}~\cite{Lastfm}, which are widely used for evaluating music recommendation algorithms, to analyze the user behavior of requesting songs. The reason why we chose these two music datasets is that a song could be requested by a user many times, such that the ground truth of user preference can be obtained from the frequency of each user's requests for each song.
To capture the main trend of user demand statistics, we choose the $100$ most active users and the 500 most popular files requested by these users for analysis, which account for more than 90\% of the requests in the datasets.

\begin{figure}[!htb]
	\centering
	\subfigure[(Global) content popularity]{
		\label{fig:popularity} 
		\includegraphics[width=0.23\textwidth]{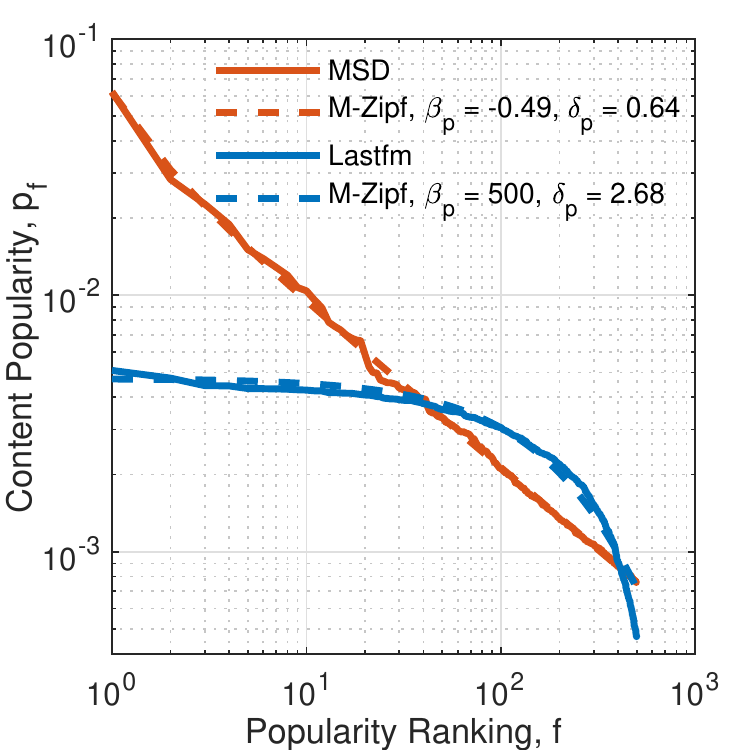} } \hspace{-4mm}
	\subfigure[User activity level]{
		\label{fig:activelevel} 
		\includegraphics[width=0.23\textwidth]{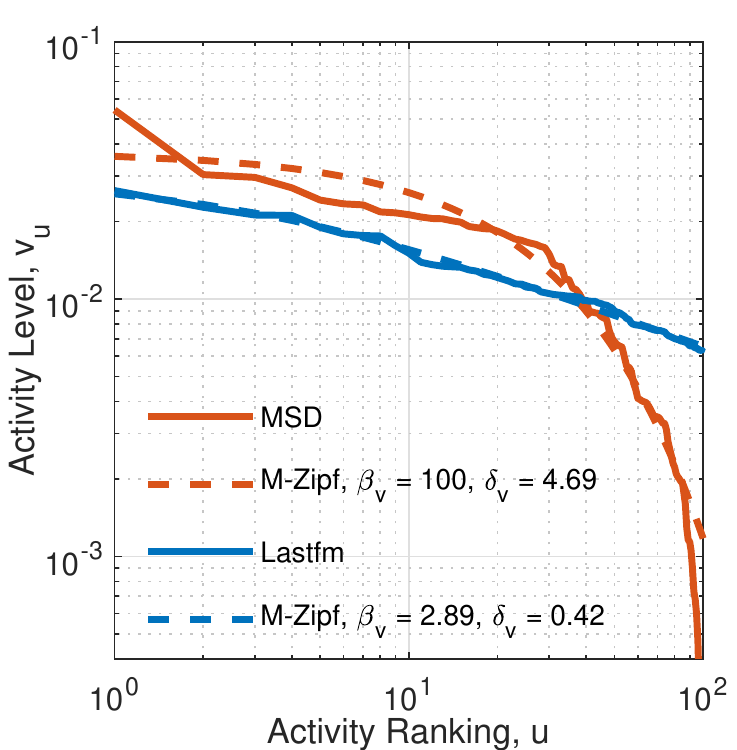} }
		\subfigure[User preference]{
		\label{fig:preference} 
		\includegraphics[width=0.23\textwidth]{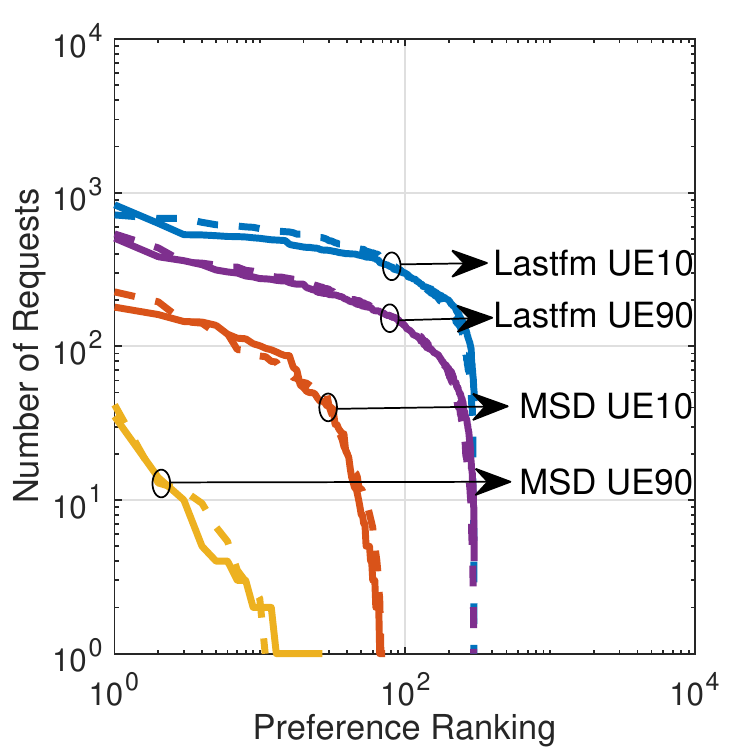}  }\hspace{-4mm}
	\subfigure[ CDF of $\cos(\mathbf{q}_u, \mathbf{q}_m)$ ]{
		\label{fig:similarity} 
		\includegraphics[width=0.23\textwidth]{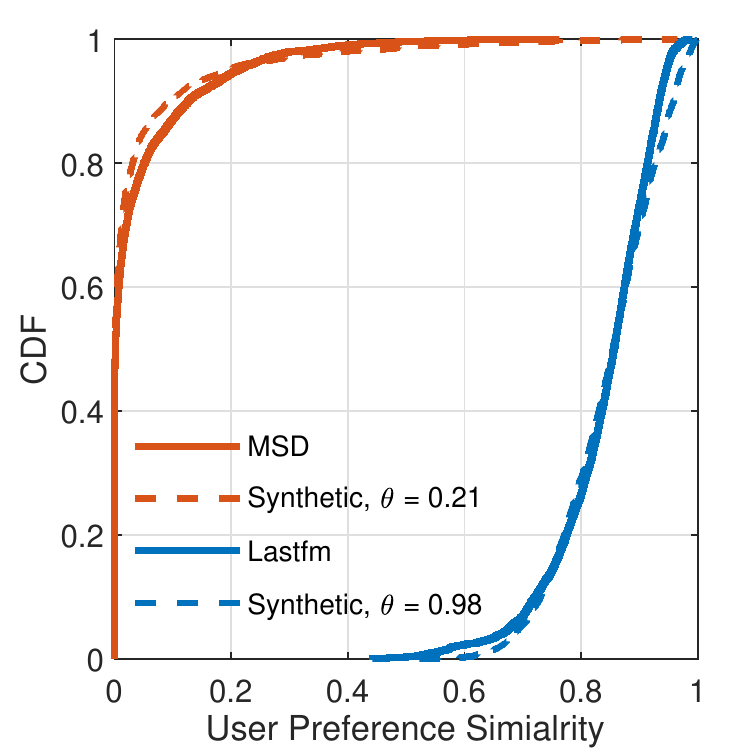}}
	\subfigure[Relation between $\theta$ and  ${\rm sim}(\mathbf Q)$ ]{
		\label{fig:theta} 
	\includegraphics[width=0.27\textwidth]{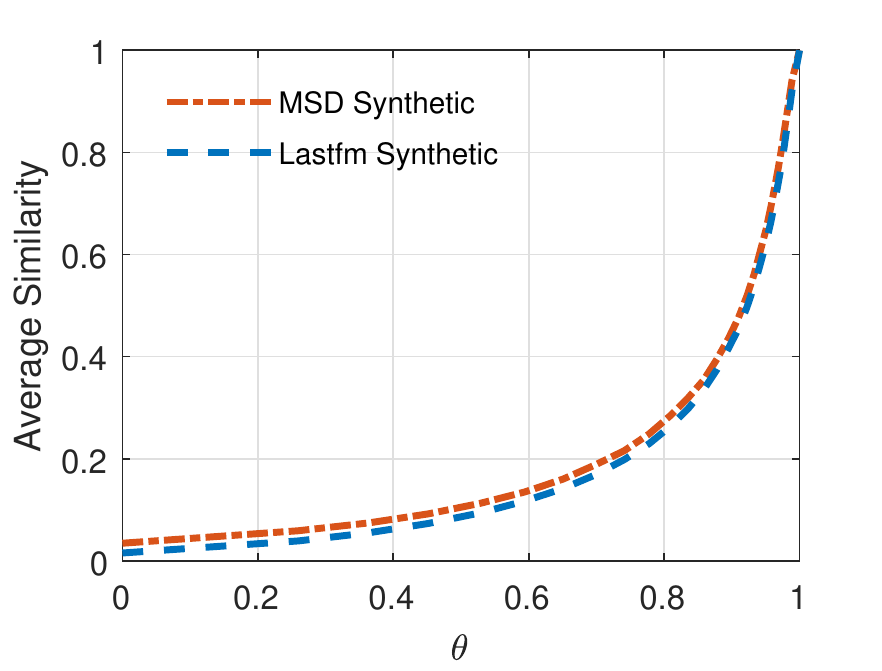}}
	\caption{Analysis of MSD and Lastfm datasets. For providing more clear figures, we show the number of requests instead of request probability in (c). The dashed lines in (c)-(e) are plotted from the synthetic user preference in Section IV-B. }
	\label{fig:pa} 
\end{figure}

In Fig. \ref{fig:popularity}, we show the content popularity in descending order. We can see that the popularity for both datasets can be fitted as Mandelbrot-Zipf distribution (M-Zipf) with expression $p_f = (f + \beta_p)^{-\delta_p}/\sum_{n=1}^{N_f}(n + \beta_p)^{-\delta_p}$, where $\delta_p$ is the Zipf skewness parameter and  $\beta_p$ is the plateau factor~\cite{Individual}. When $\beta_p=0$, it degenerates into Zipf distribution. The fitted M-Zipf parameters  are $\beta_p = 500$, $\delta_p = 2.68$ for Lastfm, and  $\beta_p = -0.49$, $\delta_p = 0.64$ for MSD, respectively.

In Fig. \ref{fig:activelevel}, we show the activity level in descending order. We can see that user activity level distribution can also be well fitted as M-Zipf distribution for both datasets, whose parameters are $\beta_v = 2.89$, $\delta_v = 0.42$ for Lastfm, and $\beta_v = 100$, $\delta_v = 4.69$ for MSD, respectively.

In Fig. \ref{fig:preference}, we show the number of requests for each file of the $10$th (i.e., more active) and $90$th (i.e., less active) users in the two datasets. To show the shape of user preference, we re-rank the files by the number of requests for each user. We can see that user preference is also skewed. Besides, it can be fitted by M-Zipf distribution  (not shown to make the figure more clear) rather than Zipf distribution as assumed in~\cite{liujuan}, and the parameters are quite different for users. In fact, the preference ranking is also different for each user. For example, in MSD, the top-$5$ preferable files of UE$_{10}$ and UE$_{90}$ are files with popularity ranking $[52,36,41,20,31]$ and files with ranking $[186,50,24,97,158]$, respectively.

In Fig. \ref{fig:similarity}, we show the cumulative distribution function (CDF) of the cosine similarity between every two-user pair. We can see that the results are quite different for Lastfm and MSD. For Lastfm, more than 90\% of the user preference similarity is larger than $0.8$, while for MSD, about 80\% of similarity is smaller than $0.2$. The average  similarity can be computed by~\eqref{eqn:avecos}, which are $0.84$ for Lastfm and  $0.04$ for MSD, respectively.

These two datesets illustrate that content popularity, activity level distribution, user preference and preference similarity vary significantly for different catalogs of contents and different groups of users. While the obtained specific distribution and parameters are non-generic, these results can demonstrate that the assumptions on identical user preference and activity level are untrue.


\subsection{User Preference Synthesization}
 Any real dataset is generated by a specific collection of users for a specific catalog of contents. It is impossible to evaluate caching policies with all real datasets. This calls for a synthetic method to generate data that can reflect user behavior in requesting contents with flexibly controlled key factors affecting caching, i.e., content popularity, user preference, preference similarity and activity level.
If we synthesize user preferences using the methods in~\cite{liujuan,Individual} and aggregate them to obtain content popularity, the content popularity will change when we adjust user preference similarity or activity level, then we cannot differentiate the impact of each factor.

In what follows, we provide an algorithm to synthesize user preferences with adjustable average similarity for any given content popularity and user activity level. The basic idea of the algorithm is as follows. (i) To synthesize generic data without specifying the distribution, we only consider the relation between user preference and content popularity and the probability constraint. The distribution (and hence the shape) of user preference is implicitly determined by the distribution of content popularity and the similarity among user preferences. In particular, from the relation $\sum_{u=1}^{N_u} v_u q_{uf} = p_f$ in \eqref{eqn:relation} and the probability constraint $\sum_{f=1}^{N_f} q_{uf} = 1$ with $0\leq q_{uf}\leq 1$, we can find an upper bound of the preference of the $u$th user for the $f$th file as $\bar q_{uf} = \min\{\frac{p_f}{v_u},1\}$. (ii) Considering that content popularity is the average of user preferences, a small deviation from the content popularity indicates a large user preference similarity. This suggests that we can control the similarity by introducing a parameter $\theta$ to adjust the variance of $q_{uf}$, given that directly controlling user preference similarity according to \eqref{eqn:avecos} is hard. In particular, we can use $\theta$  ($0\leq \theta \leq 1$) to adjust the variance by selecting $q_{uf}$ uniformly from $[\theta p_f, \theta p_f + (1 - \theta)\bar q_{uf}]$, e.g., when $\theta = 1$,  $q_{uf} = p_f$; when $\theta = 0$, the variance of $q_{uf}$ achieves the maximal value. (iii) To ensure the selected random variable satisfying the relation in \eqref{eqn:relation} and the probability constraint, the preference of each user is generated in a successive manner.

In particular, We first randomly chose a user and determine its preference. For the $u$th user, we randomly choose a file $f_1$ from the file set $\mathcal{F} = \{1, \cdots, N_f\}$ and determine $q_{uf_1}$. Then, we remove $f_1$ from $\mathcal{F}$. Because $\sum_{f=1}^{N_f} q_{uf} = 1$, the sum of the preferences of the $u$th user for the files in $\mathcal{F}$, denoted as $l \triangleq \sum_{f\in \mathcal{F}} q_{uf}$, should be updated as $l = 1 - q_{uf_1}$. Again, we randomly choose a file $f_2$ from $\mathcal{F}$. Note that  $q_{uf_2}$ is now upper bounded by $\bar q_{uf_2} = \min\{\frac{p_{f_2}}{v_u}, l\}$. After randomly setting $q_{uf_2} \in [\theta p_{f_2}, \theta p_{f_2} + (1 - \theta)\bar q_{uf_2}]$, $q_{uf_2}$ is further adjusted as $q_{uf_2} \leftarrow \min\big\{  q_{uf_2}\big(\frac{l}{\sum_{f'\in \mathcal{F}} p_{f'} }\big){}^{\theta}, \bar q_{uf_2}\big\}$ by a scaling factor $\big(\frac{l}{\sum_{f'\in \mathcal{F}} p_{f'} }\big){}^{\theta}$ to control the deviation of the mean value of user preference from content popularity, where  $\sum_{f'\in \mathcal{F}} p_{f'}$ is the sum of content popularity of the files in $\mathcal{F}$. When user preference is identical to content popularity, i.e., $q_{uf} = p_f$ for $\forall f$, we have $l =  \sum_{f'\in \mathcal{F}} p_{f'} $. When $l>\sum_{f'\in \mathcal{F}} p_{f'} $ (i.e., the mean value of  user preference for files in $\mathcal{F}$ exceeds the content popularity), the factor $\big(\frac{l}{\sum_{f'\in \mathcal{F}} p_{f'} }\big){}^{\theta} \geq 1$, which increases the value of $q_{uf_2}$ and decrease the value of $l = 1 - q_{uf_1} - q_{uf_2}$ to make $l$ closer to $\sum_{f'\in \mathcal{F}} p_{f'}$  in the next iteration. This reduces the deviation of mean value of user preference from content popularity, and vice versa. The exponent $\theta$ controls not only the variance of user preference as we mentioned before, but also the scaling factor here, which controls the mean value of user preference.
As $\theta$ increases, the scaling factor makes the mean value of user preference closer to the content popularity.
By repeating the procedure for the rest of files in $\mathcal{F}$,  ${\mathbf q}_u$ can be obtained. Next, we update the content popularity for the rest of users by subtracting $v_u\mathbf{q}_u$ from the original content popularity, and continue similar procedure. Finally, $\mathbf{Q}$ can be obtained. The whole procedure of the synthesization is shown in Algorithm~2.

\begin{algorithm}
\caption{Synthesize user preference $\mathbf{Q}$ for given $\mathbf{p}$ and $\mathbf v$}
\begin{algorithmic}[1]
	\REQUIRE $\mathbf{p}$, $\mathbf v$, $\theta$
	\ENSURE $\mathbf Q$
	\STATE $\mathbf Q \leftarrow \mathbf 0$, $\boldsymbol{\rho} \leftarrow \mathbf p$, $\mathcal{U}\leftarrow \{1, \cdots, N_u\}$   
	\WHILE{$\mathcal{U}$ is not empty}
		\STATE Randomly chose a user $u$ in $\mathcal{U}$
		\STATE $\mathcal{F} = \{1, \cdots, N_f\}$, $l \triangleq \sum_{f\in\mathcal{F}}q_{uf} = 1$
		\WHILE{$l>0 $ and $\mathcal{F}$ is not empty}
			\STATE Randomly chose a file $f$ in $\mathcal{F}$
			\STATE $\bar{q}_{uf} \leftarrow \min\left\{\frac{\rho_f}{v_u},l\right\}$, for $f = 1, \cdots, N_f$
			\STATE Set $q_{uf}\in [\theta p_f, \theta p_f + (1-\theta)\bar{q}_{uf}]$ randomly
			\STATE $q_{uf} \leftarrow \min\Big\{  q_{uf}\big(\frac{l}{\sum_{f'\in \mathcal{F}} p_{f'} }\big){}^{\theta}, \bar q_{uf}\Big\}$
			\STATE $l\leftarrow l - q_{uf}$, remove $f$ from $\mathcal{F}$
		\ENDWHILE
		\IF{$l>0$}
			\STATE $ \tilde{\mathcal{F}} \leftarrow \{ f ~ | ~q_{uf} < \bar q_{uf}\}$
			\WHILE{$l > 0$}
				\STATE Randomly chose a file $f'$ in $\tilde{\mathcal{F}}$
				\STATE $\tilde q_{uf'}\leftarrow \min\{q_{uf'} + l, \bar q_{uf'}\}$
				\STATE $l \leftarrow l - (\tilde q_{uf'} - q_{uf'})$, $q_{uf'} \leftarrow \tilde q_{uf'}$
				\IF{$q_{uf'} = \bar q_{uf'}$}
					\STATE Remove $f'$ from $\tilde{\mathcal{F}}$
				\ENDIF
			\ENDWHILE
		\ENDIF
		\STATE $\boldsymbol{\rho} \leftarrow \boldsymbol{\rho} \!-\! v_u \mathbf{q}_u$,
		 $\mathbf p \leftarrow \frac{\boldsymbol{\rho}}{\sum_{f=1}^{N_f}  \rho_f}$, remove $u$ from $\mathcal{U}$
	\ENDWHILE
\end{algorithmic}
\end{algorithm}

The relation between $\theta$ and average  similarity of user preference for the two datasets is shown in Fig.~\ref{fig:theta}, from which we can see that the average similarity increases monotonically with $\theta$ as expected. When generating synthetic data, we can obtain $\theta$ with desired average similarity from Fig.~\ref{fig:theta} and then use the obtained $\theta$ together with $\mathbf p$ and $\mathbf v$ to synthesize user preference.

In Fig. \ref{fig:preference}, we plot the synthetic user preferences of the $10$th and $90$th active users based on the popularity and activity level of Lastfm and MSD datasets. We chose $\theta = 0.98$ and $\theta = 0.21$ for Lastfm and MSD, respectively, so that the average similarities of user preferences in the real datasets and synthetic data are identical. We can see that the distribution of the synthetic user preference is almost the same as the datasets.

In Fig. \ref{fig:similarity}, we further plot the CDF of the similarity $\cos(\mathbf{q}_u, \mathbf{q}_m)$ of the synthetic user preference to compare the distribution of user preference similarity between the synthetic data and real data. We can see that the synthetic data can fit  both real datasets well.

Both Algorithm~2 and the method in~\cite{Individual} can fit real datasets well, but they have their own pros and cons, leading to different targeting applications. In~\cite{Individual}, the user preference for each content is synthesized as the probability that a user prefers a specific genre multiplied by the popularity of a content within this genre. Such model is able to capture some inner structure of the data, which can help understanding user's request pattern and is more flexible in controlling the user preference with more parameters. On the other hand, Algorithm~2 does not require the direct modeling of user preference, which may avoid the bias introduced by particular datasets. Moreover, Algorithm~2 is able to control user preference similarity, activity level and content popularity separately, which can differentiate the impact of each factor for performance evaluation. Besides, in some scenarios, user preference may exhibit a clustering effect such that users in one cell have strong correlation in their preferences. Algorithm~2 is able to capture such clustering effect by implementing in a cascading fashion as follows. We can first generate the preferences of $M$ virtual users as cluster centers by Algorithm~2 and use $\theta$ to adjust the distance (measured by similarity) between the cluster centers. Then, we can generate user preference within each cluster, again using Algorithm~2 by regarding the preference of virtual user as the content popularity within the cluster and use $\theta$ to adjust the user preference similarity inside each cluster. With the generated user preferences, we can assign the same location probability distribution $\mathbf a_u$ to the users within the same cluster so that users in one cell will have strong correlation in their preferences.

\section{Simulation Results}
In this section, we compare the performance of the proposed caching policies with prior works, and analyze the impact of user preference similarity, user activity level and spatial locality by simulation based on the synthetic data.

Consider $N_b = 7$ cells with radius $D = 40$ m as shown in Fig.~\ref{fig:layout}.\footnote{When the BSs are located randomly as in Fig.~\ref{fig:layout_general}, the performance is similar and hence not provided due to space limitation.} The backhaul bandwidth and the downlink transmission bandwidth for each user are set as $C_u^{{\rm bh}} = 1$ Mbps and $W_u = 5$ MHz, respectively. We consider Rayleigh fading channels and pathloss modeled as $35.5+37.6\log_{10}(r_{ub})$ in dB. The transmit power of BS and the noise power are $21$~dBm and $-174$~dBm/Hz, respectively. To reduce simulation time, the total number of users in the considered region is set as $N_u = 50$, $N_f = 100$ files, and  $N_c = 10$, i.e., each BS can cache 10\% of the total files. The Zipf's skewness parameters for global content popularity  and activity level distribution are set as $\delta_p = 0.6$ and $\delta_v = 0.4$, respectively, according to the data analysis in Section \ref{subsec:a}. The user location probability distribution is modeled as Zipf distribution with skewness parameter $\delta_a = 1$ based on the measured data in~\cite{traffic,traffic2}. A larger value of $\delta_a$ indicates that a user sends requests in few cells with high probability. Unless otherwise specified, this setting is used throughout the simulation.

The following caching policies are simulated for comparison:
\begin{enumerate}
	\item {\em``Local Pop"}: This is the caching policy used in~\cite{ahlehagh2014video,bigdata}, where each BS caches the most popular files according to the local content popularity within its cell given by \eqref{eqn:local}.
	\item {\em ``Femtocaching"}: This is a deterministic caching policy based on  global content popularity, identical activity level and known user location, which is optimized under the same assumptions as the policy proposed in~\cite{femtocachingTIT}. For a fair comparison, we obtain this policy to maximize the network utility. To show the impact of  location uncertainty, the policy is obtained based on one realization of user location and remains unchanged for other realizations.
	\item {\em``Femtocaching (UP)"}: This is a deterministic caching policy based on user preference, activity level and known user location, which is optimized under the same assumptions as a policy proposed in~\cite{liujuan} except using the network utility as objective function and considering heterogeneous activity level. Again, the policy is obtained based on one realization of user location and remains unchanged for other realizations. The only difference of this policy with Policy~1 lies in the assumption on user location, since Policy~1 degenerates into deterministic policy as stated in Corollary 1.
	\item {\em``Policy~1 (Pop)"}: This is a probabilistic caching policy obtained from problem ${\sf P}_1$ by using global content popularity (i.e., replacing $q_{uf}$ with $p_f$) and setting identical user activity level. The only difference of this policy with ``Femtocaching" lies in the assumption on user location, according to Corollary 1.
	\item {\em``Policy~2 (Pop)"}: This is a probabilistic caching policy obtained from problem ${\sf P}_2$ by using global content popularity and setting identical user activity level.
\end{enumerate}

\begin{figure}[!htb]
	\centering	
	\subfigure[Network averge rate]{
		\label{fig:R_vs_cos} 
		\includegraphics[width=0.4\textwidth]{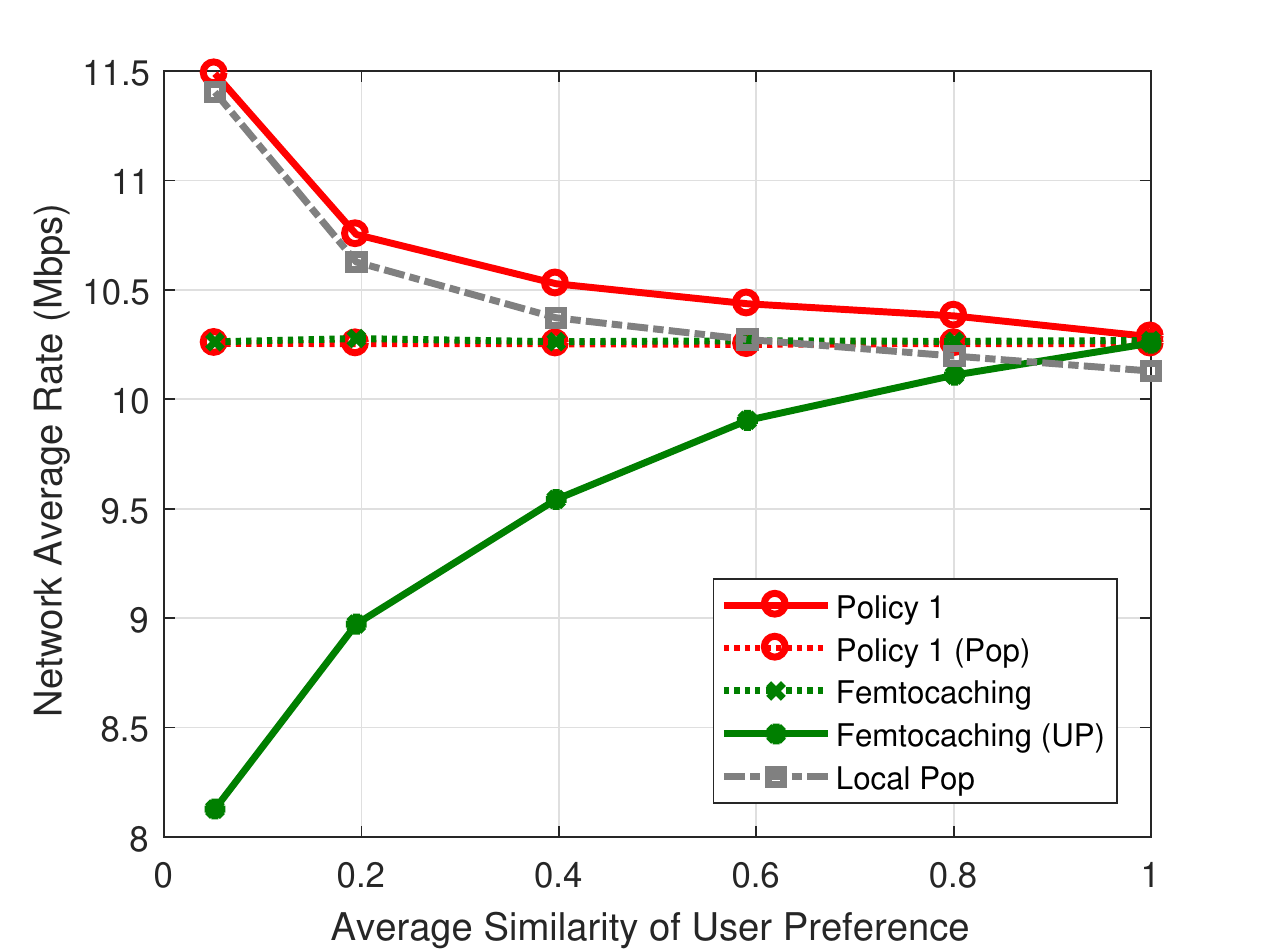}}
	\subfigure[Minimal average rate]{
		\label{fig:min_vs_cos} 
		\includegraphics[width=0.4\textwidth]{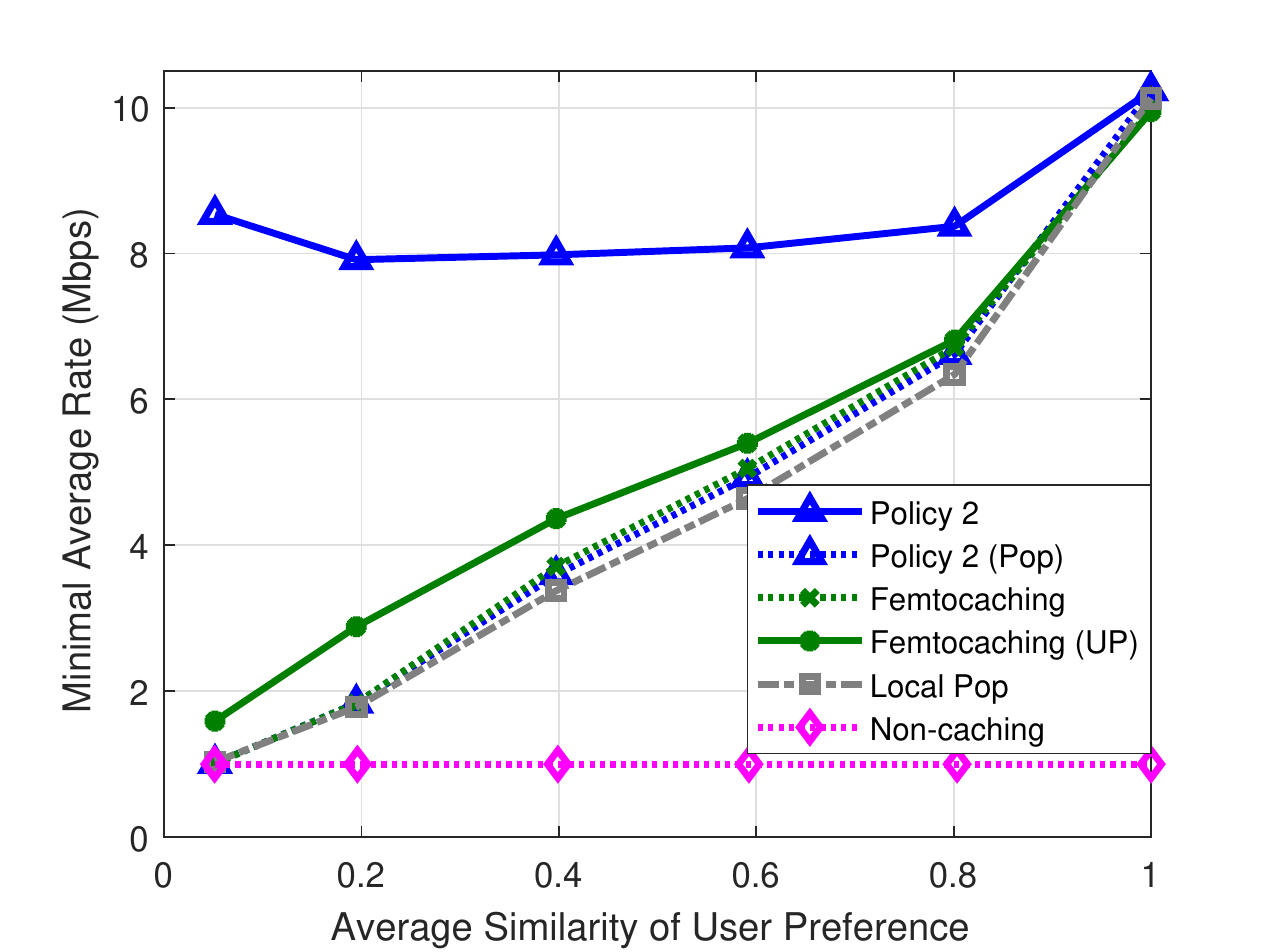}}
	\caption{Impact of user preference similarity.}
	\label{fig:Q}
\end{figure}

In Fig.~\ref{fig:R_vs_cos}, we show the impact of user preference
similarity on the network performance.
 We can see that ``Policy~1 (Pop)" almost performs the same as  ``Femtocaching". However, ``Femtocaching (UP)" is inferior to ``Femtocaching" when user preference is less similar, because ``Femtocaching (UP)" does not consider the uncertainty of user location. These results indicate that location uncertainty has large impact when user preference is heterogeneous. The network average rate of Policy~1 is the highest and the gain over global popularity based methods (i.e., ``Femtocaching" and ``Policy~1 (Pop)") increases with the decrease of preference similarity. This comes from the file diversity and more skewed user preference as explained in Section~III-C.

In Fig. \ref{fig:min_vs_cos}, we show the impact of user preference similarity on user fairness. We can see that Policy~2 provides higher minimal average rate than other baseline policies unless ${\rm sim}(\mathbf Q)=1$. However, ``Policy~2 (Pop)"  is inferior to  ``Femtocaching (UP)", because the knowledge of user preference is important for improving user fairness. The minimal average rate of Policy~2 first decreases and then increases with the preference similarity, due to the combination effects of heterogeneous user preference as explained in the end of Section III-C. We also show the performance of a corresponding non-caching system, over which Policy~2 can provide 800\%~$\sim$~10000\% gain in terms of the minimal average rate. Since the network average rate and minimal average rate of the non-caching system are almost the same (that is about $1$ Mbps in the considered setting), which are not affected by user preference similarity, spatial locality and activity level skewness parameters, we do not show the related results in the sequel.

\begin{figure}[!htb]
	\centering
	\subfigure[Network averge rate]{
		\label{fig:R_vs_a} 
		\includegraphics[width=0.4\textwidth]{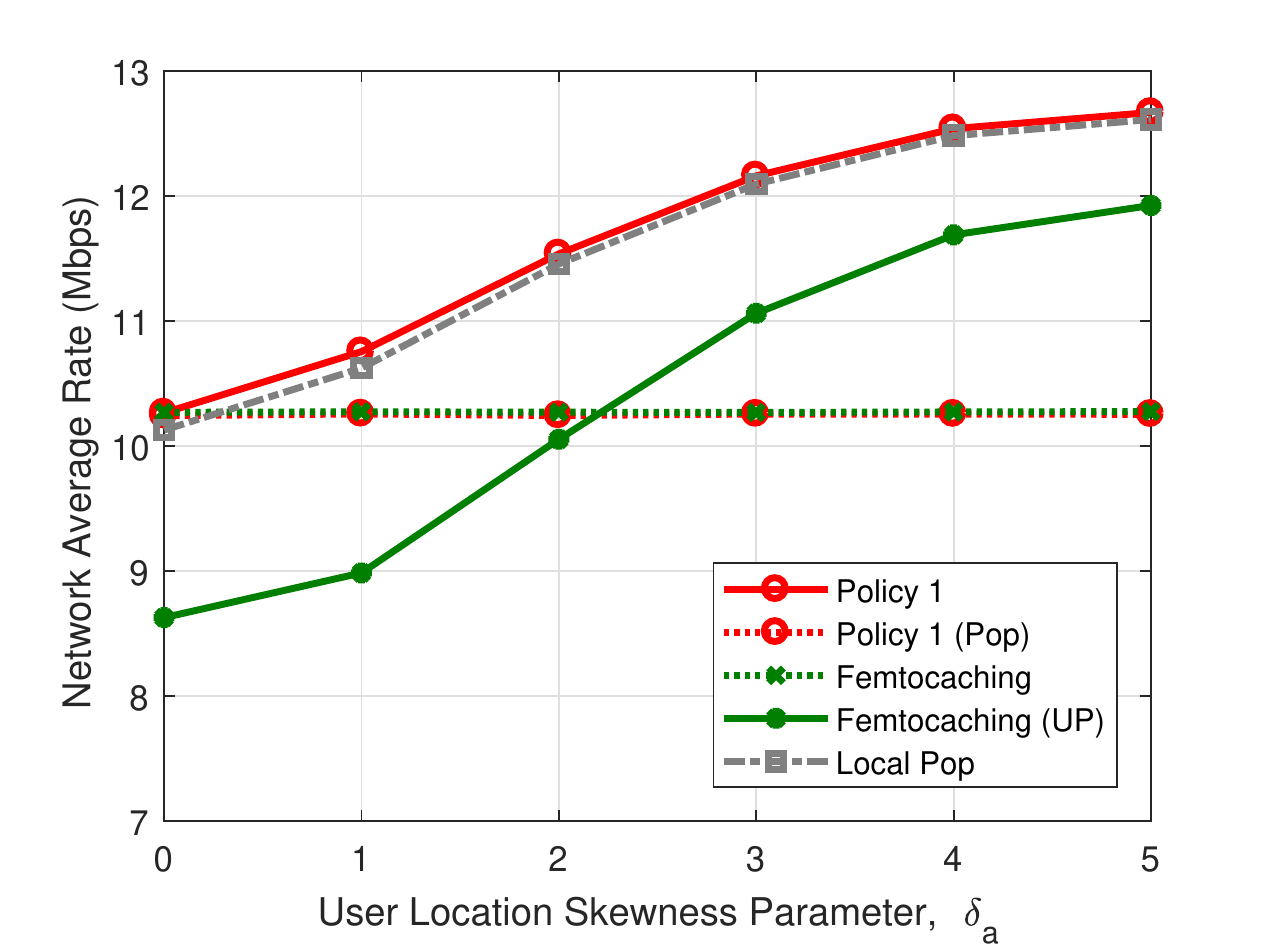}}
	\subfigure[Minimal average rate]{
		\label{fig:min_vs_a} 
		\includegraphics[width=0.4\textwidth]{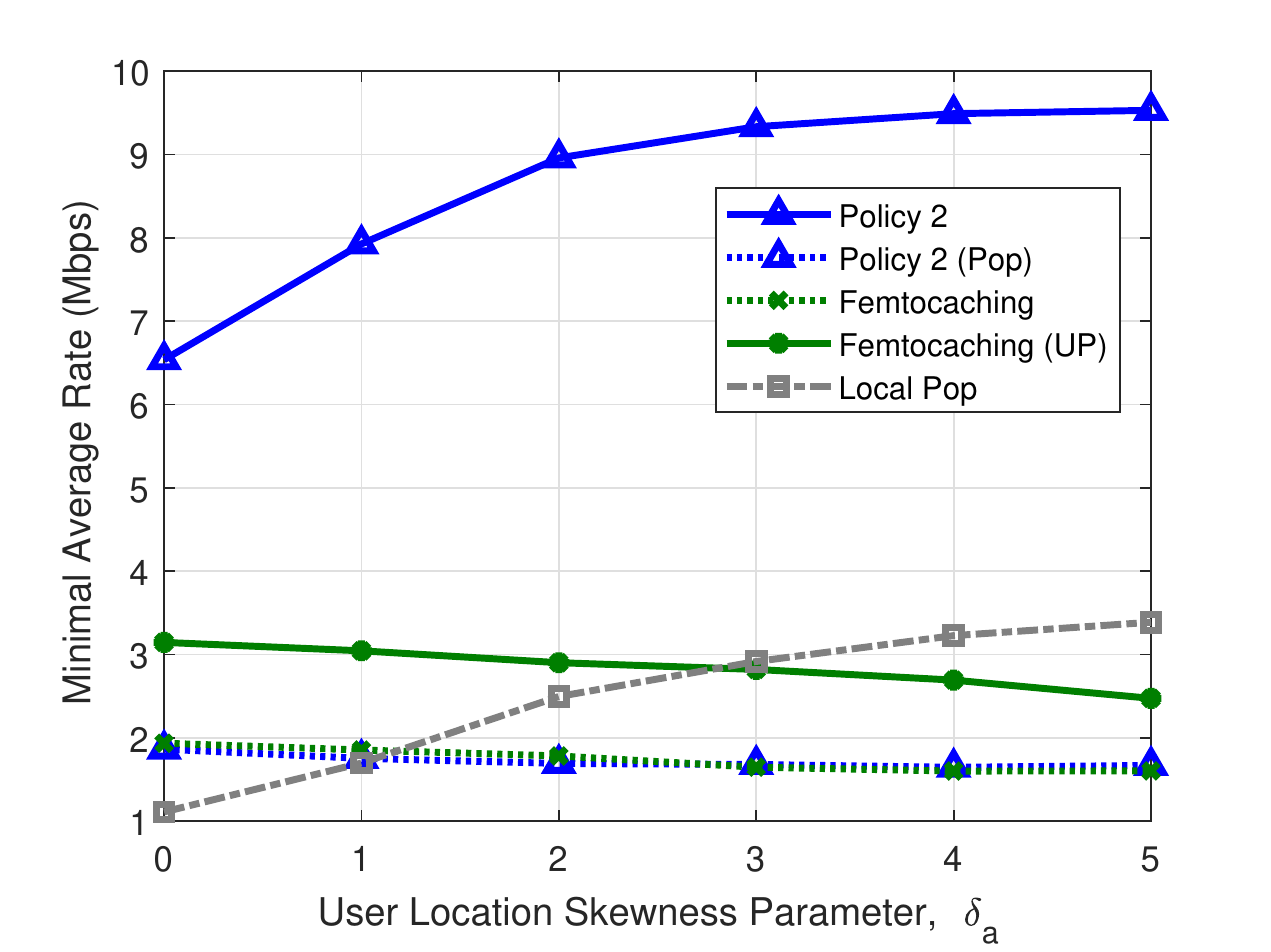}}
	\caption{Impact of user location skewness parameter, $\delta_a$. The average user preference similarity is set as ${\rm sim}(\mathbf{Q}) = 0.2$.}
	\label{fig:a} 
\end{figure}

In Fig.~\ref{fig:a}, we show the impact of spatial locatity. As shown in Fig.~\ref{fig:R_vs_a}, the network average rates of ``Local Pop" and user preference based caching policies increase with $\delta_a$. When $\delta_a  = 0$, i.e., each user sends requests in each of the $N_b$ cells  with equal probability, ``Policy~1 (Pop)" achieves the same performance as Policy~1, which verifies Corollary 2. When $\delta_a  = 5$, i.e., each user is with $ \frac{1^{-5}}{\sum_{i=1}^{7}i^{-5}} = 0.96$ probability located in the most probable cell, Policy~1 has 30\% gain over ``Policy~1 (Pop)". This suggests that the gain of network average rate by exploiting user preference highly relies on the spatial locality of user. As shown in Fig. \ref{fig:min_vs_a}, the minimal average rates of ``Local Pop" and Policy~2 also increase with $\delta_a$. When $\delta_a = 5$, Policy~2 can triple the minimal average rate compared with other caching policies. On the contrary, spatial locality has little impact on both network average rates and minimal average rates achieved by the global popularity based caching policies (i.e., ``Femtocaching" and ``Policy~1 (Pop)").

\begin{figure}[!htb]
	\centering
	\subfigure[Network averge rate]{
		\label{fig:suc_vs_s} 
		\includegraphics[width=0.4\textwidth]{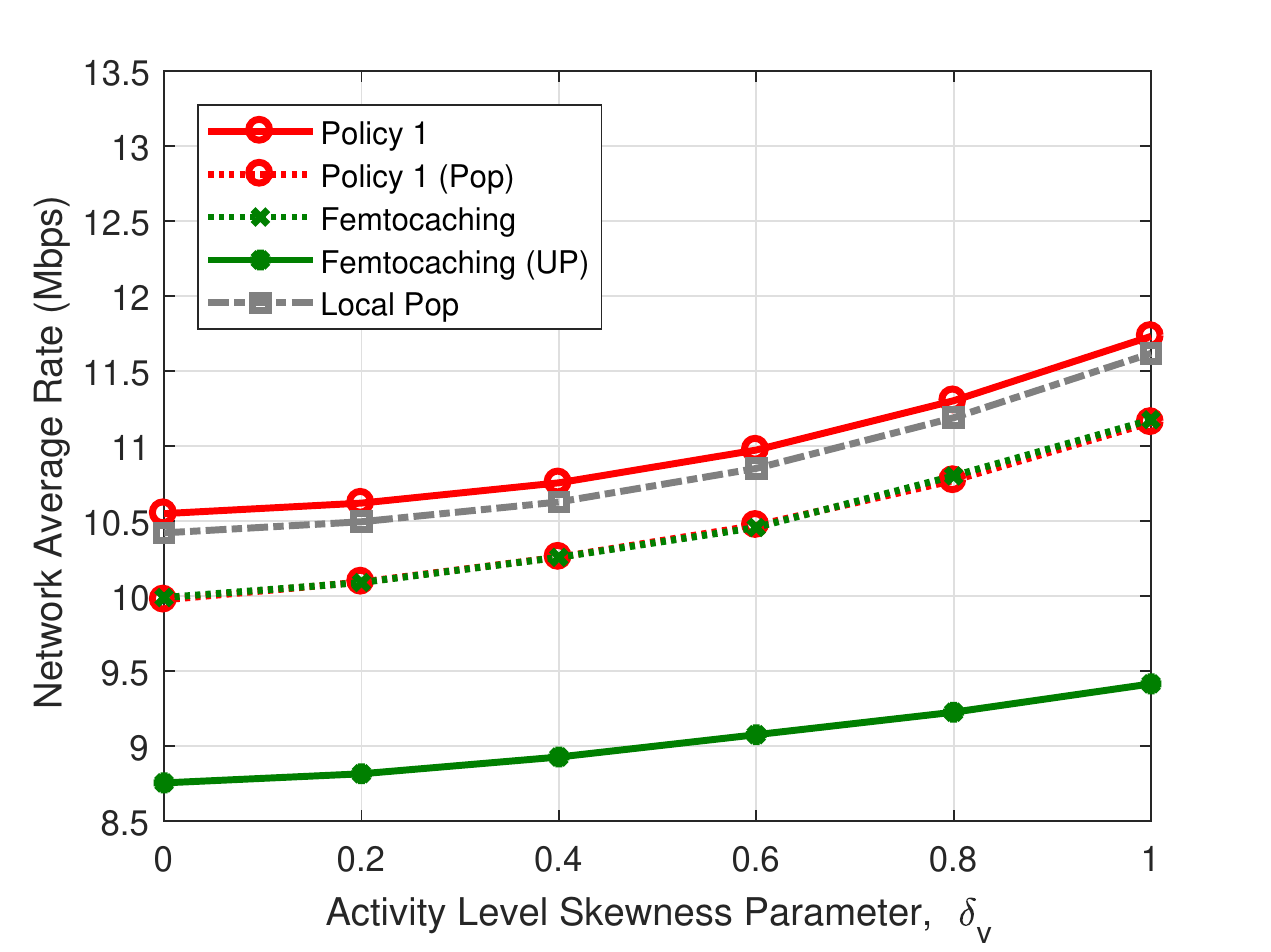}}
	\subfigure[Minimal average rate]{
		\label{fig:min_vs_s} 
		\includegraphics[width=0.4\textwidth]{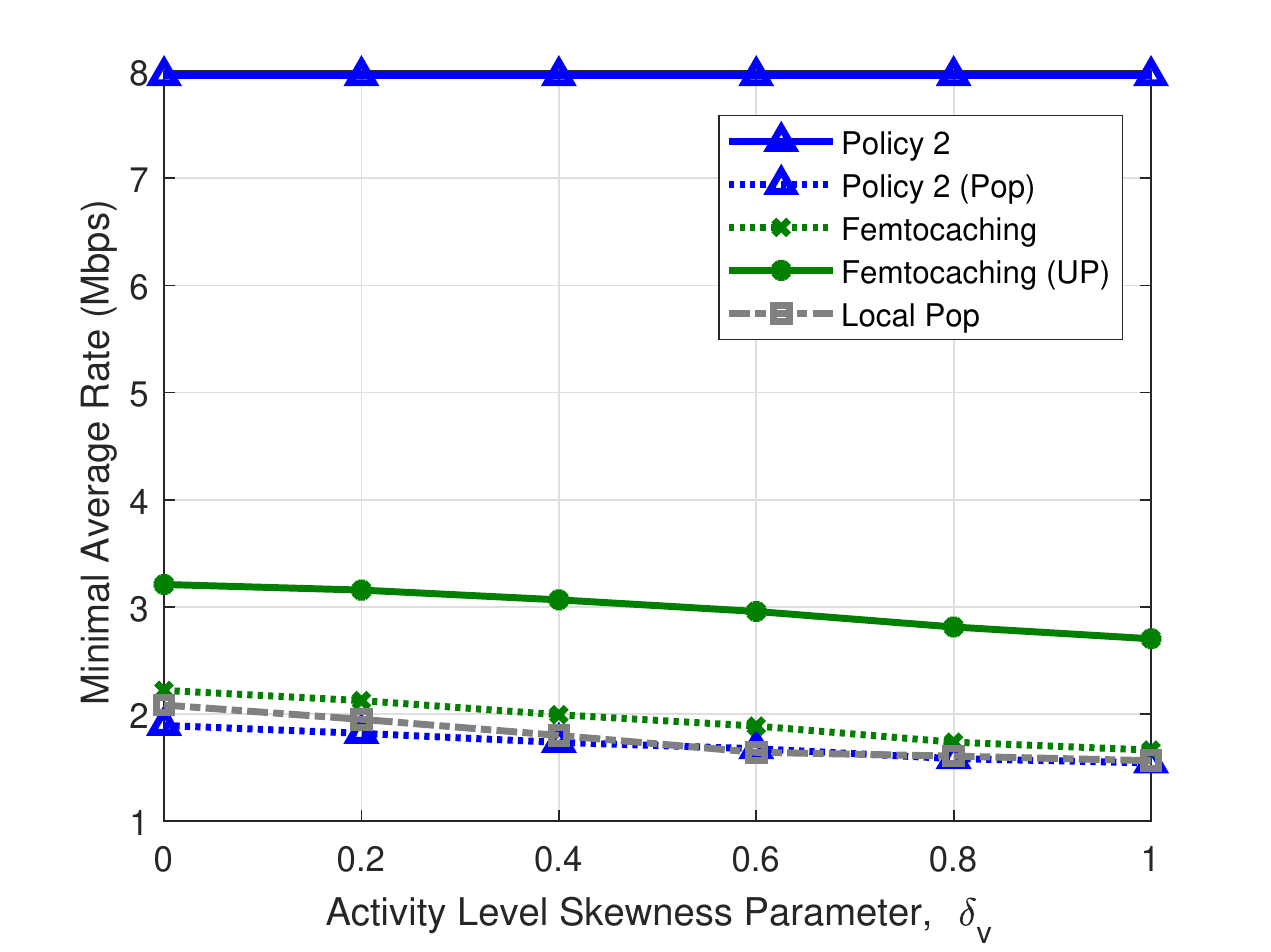}}
	\caption{Impact of user activity level skewness parameter, $\delta_v$. The average user preference similarity is set as ${\rm sim}(\mathbf{Q}) = 0.2$.}
	\label{fig:s} 
\end{figure}

In Fig. \ref{fig:s}, we show the impact of user activity level skewness.  As shown in Fig. \ref{fig:suc_vs_s}, the network average rates of all caching policies increase with $\delta_v$. This is because when the user activity level distribution is more skewed, the caching solutions are determined more by the preferences of highly active users. As a result, the average rates of these active users increase, which yields higher network average rate. As shown in Fig. \ref{fig:min_vs_s}, the minimal average rate of Policy~2 is highest, and the performance gain increases with $\delta_v$.

In Figs. \ref{fig:Q}, \ref{fig:a} and \ref{fig:s}, ``Femtocaching" and ``Policy~1 (Pop)" almost perform the same, and ``Femtocaching" and ``Policy~2 (Pop)"  perform closely. This is because when user preference and activity level are regarded as identical, there is no difference among users statistically as explained in~\cite{femtocachingTIT}, and hence the uncertainty of user location has little impact.

As shown in previous results with average achievable rate as the network utility, the performance of ``Local Pop" and Policy~1 is very close. This can be explained by Corollary 3 since the average rate when downloading from the nearest BS's cache is much larger than the average rate when downloading from the second nearest BS's cache. When using success probability as the network utility, the performance trends are similar, but the gap between Policy~1 and ``Local Pop" is larger (not shown due to the lack of space). This is because the success probabilities when the user downloads from the nearest and second nearest BSs' caches will be close if $\gamma_0$ is relatively small, e.g., $\gamma_0 = -5$ dB (equivalent to $2$ Mbps rate requirement with $W_u = 5$ MHz).

\begin{figure}[!htb]
\centering
\includegraphics[width=0.4\textwidth]{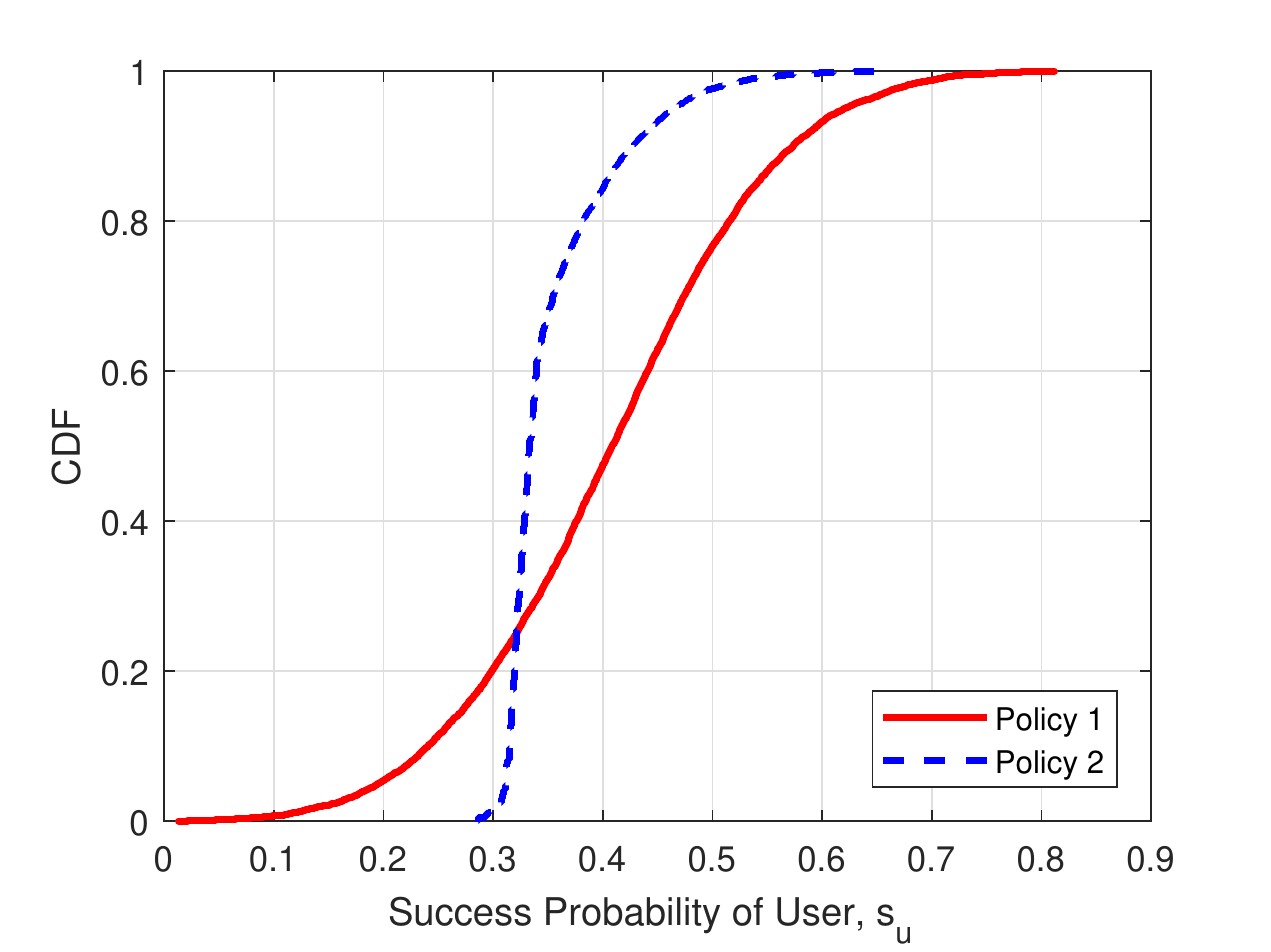}
\caption{CDF of success probability, $\gamma_0 = -5$ dB, ${\rm sim}(\mathbf{Q}) = 0.2$.}
\label{fig:cdf}
\end{figure}


In Fig.~\ref{fig:cdf}, we show the CDF of success probability achieved by Policy~1 and Policy~2. We can see that with Policy~2, the proportion of users with low success probability (e.g., $s_u < 0.3$) is lower than Policy~1, while the proportion of users with high success probability is also lower, resulting in a tradeoff between network average performance and  user fairness.

\begin{figure}[!htb]
	\centering
	\includegraphics[width=0.4\textwidth]{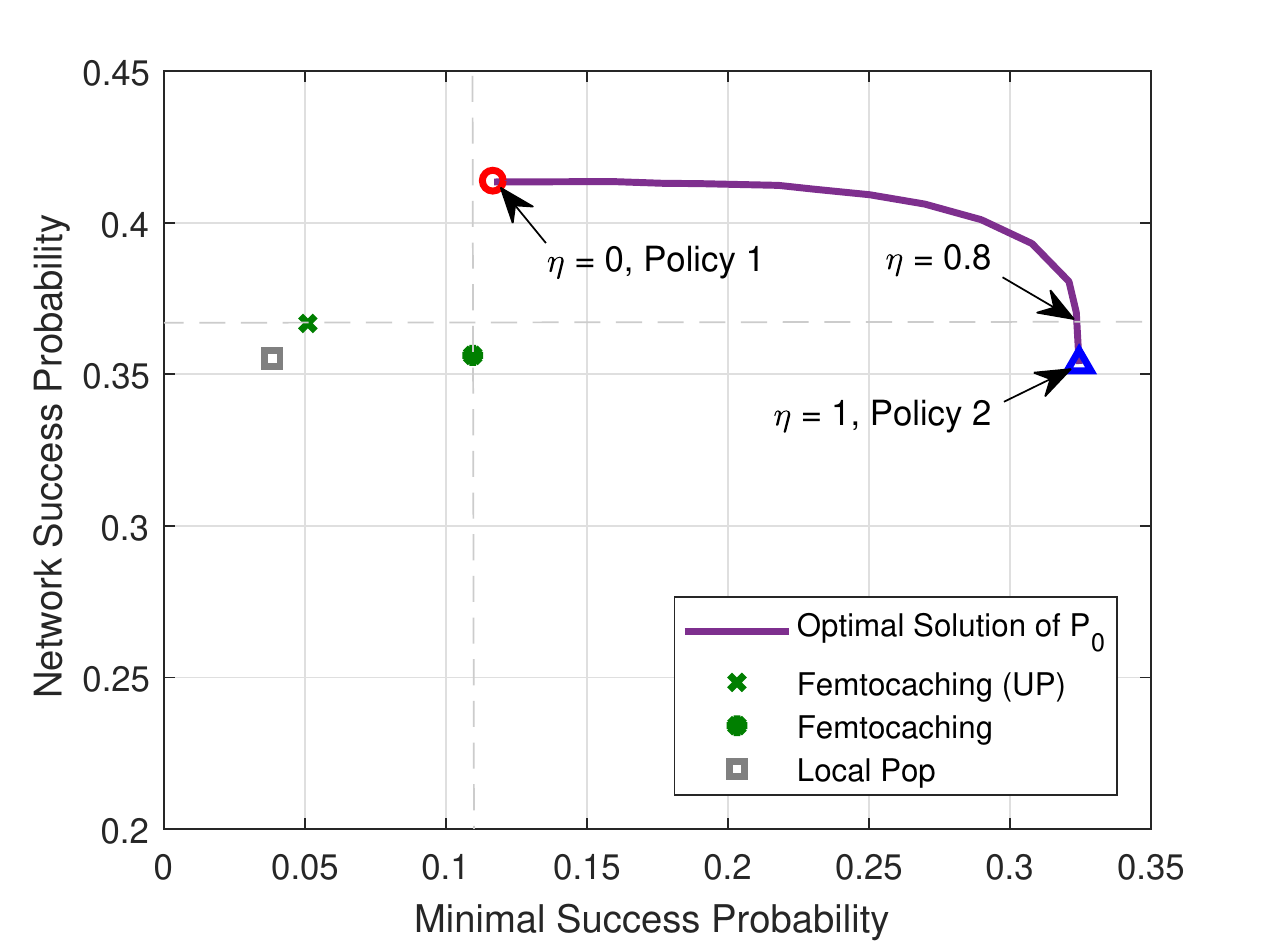}
	\caption{Tradeoff between network performance and user fairness, $\gamma_0=-5$~dB, ${\rm sim}(\mathbf{Q}) = 0.2$.}
	\label{fig:tradeoff}
\end{figure}

In Fig.~\ref{fig:tradeoff}, we show such tradeoff by solving problem ${\sf P}_0$ with different values of  $\eta$. It is shown that the proposed policy can improve network performance and user fairness simultaneously compared with the baseline policies in a wide range of $\eta$ (i.e., $0\leq \eta \leq 0.8$).

From the observation in Section II, the corollaries in Section III-B as well as the simulations in this section, we can summarize the assumptions that make the user preference based caching policies the same as the content popularity based caching policies in Table~\ref{tab:con}. Without these assumptions, exploiting individual user behavior is beneficial for local caching.

\begin{table*}[!htbp]
	\centering
	\caption{When user preference based caching policies will not be beneficial}
	\begin{tabular}{lll}
		\toprule
		 Objectives & Assumptions & Equivalent to caching policies based on \\
		\midrule
		\multicolumn{1}{l}{\multirow{4}[0]{*}{Network Performance}} & Identical transmission resouce \&  & \multirow{2}[0]{*}{Global/local content popularity} \\
		\multicolumn{1}{l}{} & Without spatial locality &  \\ \cline{2-3}\noalign{\smallskip}
		\multicolumn{1}{l}{} & Homogeneous user preference & Global/local content popularity \\ \cline{2-3}\noalign{\smallskip}
		\multicolumn{1}{l}{} & Identical transmission resource & Local content popularity \\ 
\cline{1-3}\noalign{\smallskip}
		\multirow{3}[0]{*}{User Fairness} & Identical transmission resource \&& \multirow{3}[0]{*}{Global/local content popularity} \\
		& Homogeneous user preference and activity level \& &  \\
		& Identical spatial location probability distribution &  \\
		\bottomrule
	\end{tabular}%
	\label{tab:con}%
\end{table*}%

\section{Conclusion and Discussion}
In this paper, we investigated when the user behavior in terms of spatial locality, heterogeneous preference and activity level impact proactive caching by establishing a caching policy optimization framework to maximize a weighted sum of the network utility and the minimal utility among users. We showed the relation between the global content popularity, local content popularity, user preference and spatial locality. To evaluate the entangled impacts of content popularity, user preference and activity level on the performance of wireless edge caching, we provided an algorithm to synthesize user preference with given content popularity, user activity level and adjustable user preference similarity, and validated it with two real datasets. We found that the gain of exploiting individual user behavior is large under realistic settings, where user preferences are less similar, user activity level distribution is skewed, and more importantly, users are with strong spatial locality. Simulation results showed that both the network performance and user fairness achieved by the proposed policy are superior to prior works based on either content popularity or user preference.

In practice, learning individual user preference is more computationally complex than learning content popularity. Moreover, content provider (CP) usually transfers data to wireless users via secure connections~\cite{r2}. Consequently, security issues could be a barrier for MNO to learn user preference or content popularity for proactive caching, or even for reactive caching. Nevertheless, new security protocols have been proposed to enable the MNOs to perform caching on encrypted requests as discussed in~\cite{barriers}. On the other hand, there is a recent trend of the convergence in managing cache by MNOs and CPs~\cite{soft}. User preference can be learned by a CP and then shared with a MNO. Alternatively, a CP can install its own caches in wireless edge~\cite{barriers} or lease the caches from a MNO~\cite{r2}, and the MNO shares the user location distribution information to the CP for optimizing caching policy toward better user experience. Such cooperation is possible since user experience can be improved significantly, which is a win-win situation.

\section{Acknowledgment}
We sincerely thank the anonymous reviewers for their insightful comments and suggestions.
\appendices
\section{Proof of Proposition 1}
\renewcommand{\theequation}{A.\arabic{equation}}
\setcounter{equation}{0}
From \eqref{eqn:su}, by taking the expectation over user request $f$, we can obtain
\begin{multline}
s_u =\sum_{f=1}^{N_f} q_{uf}\mathbb{E}_{\mathbf{x}_u} \Bigg[\sum_{k=1}^{K} \Big[c_{f,k(\mathbf x_u)} \prod_{l=1}^{k-1} (1 - c_{f,l(\mathbf{x}_u )})  \Big] \\
\times\mathbb{P}\left( \gamma_{u,k(\mathbf{x}_u)}(\mathbf{x}_u) > \gamma_0 ~|~ \mathbf{x}_u\right)  \Bigg] \label{eqn:su1}
\end{multline}
Since the $k$th nearest BS of the $u$th user $k(\mathbf{x}_u)$ depends on user location, the expectation over $\mathbf{x}_u$ cannot move into the summation over $k$. Since we have divided each cell into small regions as shown in Fig. \ref{fig:network}, the $k$th nearest BS of the $u$th user only depends on which small region of which cell the user is located in rather than the exact location $\mathbf{x}_u$. Therefore, we can denote $k_{ij}$ as the $k$th nearest BS when the user is located in the $j$th small region of the $i$th cell (i.e., $\mathcal{D}_{ij}$). Then, based on the law of total expectation, \eqref{eqn:su1} can be derived as
\begin{align}
s_u =&\sum_{f=1}^{N_f} q_{uf}\mathbb{E}_{ij}\Bigg[\mathbb{E}_{\mathbf{x}_u \in \mathcal{D}_{ij}} \Bigg[\sum_{k=1}^{K} \Big[c_{fk_{ij}} \prod_{l=1}^{k-1} (1 - c_{fl_{ij}})  \Big] \nonumber \\
& \times \mathbb{P}\left( \gamma_{uk_{ij}}(\mathbf{x}_u) > \gamma_0 ~|~ \mathbf{x}_u \right)  \Bigg]\Bigg] \nonumber \\
 \overset{(a)}{=}& \sum_{f=1}^{N_f} q_{uf}\mathbb{E}_{ij}\Bigg[\sum_{k=1}^{K} \Big[c_{fk_{ij}} \prod_{l=1}^{k-1} (1 - c_{fl_{ij}})  \Big]  \nonumber \\
&\times \mathbb{E}_{\mathbf{x}_u \in \mathcal{D}_{ij}} \left[ \mathbb{P}\left( \gamma_{uk_{ij}}(\mathbf{x}_u) > \gamma_0 ~|~ \mathbf{x}_u \right)  \right]\Bigg] \nonumber \\
 =& \sum_{f=1}^{N_f} q_{uf}\sum_{i=1}^{N_b}
\sum_{j=1}^{J_i}\frac{a_{ui} |\mathcal{D}_{ij}|}{|\mathcal{D}_{i}|}
\sum_{k=1}^{K} \Big[c_{fk_{ij}} \prod_{l=1}^{k-1} (1 - c_{fl_{ij}})  \Big] {\sf s}_{uk_{ij}}  \label{eqn:su2}
\end{align}
where $\mathbb{E}_{ij}$ denotes the expectation over $i$ and $j$, $\mathbb{E}_{\mathbf x_u\in \mathcal{D}_{ij}}$ denotes the expectation over $\mathbf{x}_u$ within $\mathcal{D}_{ij}$, step $(a)$ safely moves $\mathbb{E}_{\mathbf x_u\in \mathcal{D}_{ij}}$ into the summation over $k$ since $k_{ij}$ does not depend on $\mathbf{x}_u$ anymore given that $\mathbf x_u\in \mathcal{D}_{ij}$,  $\frac{a_{ui} |\mathcal{D}_{ij}|}{|\mathcal{D}_{i}|}$ is the probability that  $\mathbf{x}_u \in \mathcal{D}_{ij}$, and ${\sf s}_{uk_{ij}} \triangleq \mathbb{E}_{\mathbf{x}_u\in \mathcal{D}_{ij}} \left[ \mathbb{P}\left( \gamma_{uk_{ij}}(\mathbf{x}_u) > \gamma_0 ~|~ \mathbf{x}_u \right)  \right]$ is the success probability when the user is located in $\mathcal{D}_{ij}$ and downloads from its $k$th nearest BS's cache.

To obtain ${\sf s}_{uk_{ij}}$, we first derive the success probability conditioned on given user location as
\begin{align}
&\mathbb{P}\left( \gamma_{uk_{ij}}(\mathbf{x}_u) > \gamma_0\!~ |\!~ \mathbf{x}_u \right) \! = \!\mathbb{P}\left( h_{uk_{ij}} \! \geq \! \gamma_0 r_{uk_{ij}}^{\alpha} \big( I_{uk_{ij}} + \tfrac{\sigma^2}{P}\big)~\Big| ~\mathbf{x}_u \! \right) \! \nonumber\\
&\overset{(a)}{=}\! \mathbb{E}_{I_{uk_{ij}}}\!\!\left[ \exp\left(- \gamma_0 r_{uk_{ij}}^{\alpha} \big(I_{uk_{ij}} + \tfrac{\sigma^2}{P}\big)~\Big|~ \mathbf{x}_u \right) \right] \nonumber \\
& \overset{(b)}{=} e^{-\gamma_0 r_{uk_{ij}}^{\alpha}\frac{\sigma^2}{P}} \mathbb{E}_{\mathbf h_{u}}\Bigg[ \prod_{b\in \Phi_{k_{ij}}, b \neq k} \exp\left(- \gamma_0 r_{uk_{ij}}^{\alpha}   h_{ub} r_{ub}^{-\alpha} \right) \Bigg] \nonumber \\
&= e^{-\gamma_0 r_{uk_{ij}}^{\alpha}\frac{\sigma^2}{P}}\prod_{b\in \Phi_{k_{ij}}, b \neq k_{ij}} \left(1 + \gamma_0 r_{uk_{ij}}^{\alpha} r_{ub}^{-\alpha}\right)^{-1}  \label{eqn:suk}
\end{align}
where step $(a)$ is from $h_{uk_{ij}}\sim \exp(1)$ for Rayleigh fading, step $(b)$ is upon substituting $I_{uk_{ij}} = \sum_{b\in\Phi_{k_{ij}},b \neq k_{ij}} h_{ub} r_{ub}^{-\alpha}$, and the last step is because $\{h_{ub}\}$  are independently distributed and $h_{ub} \sim \exp(1)$.
Then, by averaging over user location within small region $\mathcal{D}_{ij}$, we can obtain
\begin{align}
&{\sf s}_{uk_{ij}}(\gamma_0)  = \mathbb{E}_{\mathbf{x}_u\in \mathcal{D}_{ij}} \left[ \mathbb{P}\left( \gamma_{uk_{ij}}(\mathbf{x}_u) > \gamma_0 | \mathbf{x}_u \right)  \right] \nonumber \\
&=  \iint\limits_{\mathbf{x}_u \in \mathcal{D}_{ij}} \frac{1}{|\mathcal{D}_{ij}|} \mathbb{P}\left( \gamma_{uk_{ij}}(\mathbf{x}_u) > \gamma_0 | \mathbf{x}_u \right)  {\rm d}x_{u1}{\rm d}x_{u2}  \label{eqn:sk}
\end{align}
Finally, by substituting \eqref{eqn:suk} into \eqref{eqn:sk} and then into \eqref{eqn:su2}, Proposition 1 can be proved.

\section{Proof of Proposition 2}
\renewcommand{\theequation}{B.\arabic{equation}}
\setcounter{equation}{0}
Similar to the derivation of \eqref{eqn:su2}, we can obtain
\begin{equation}
\bar R_u= \sum_{f=1}^{N_f} q_{uf}\sum_{i=1}^{N_b}
\sum_{j=1}^{J_i}\frac{a_{ui} |\mathcal{D}_{ij}|}{|\mathcal{D}_{i}|}
\sum_{k=1}^{K} \Big[c_{fk_{ij}} \prod_{l=1}^{k-1} (1 - c_{fl_{ij}})  \Big]  {\sf R}_{uk_{ij}}  \label{eqn:Ru1}
\end{equation}
where ${\sf R}_{uk_{ij}} \triangleq \mathbb{E}_{\mathbf{x}_u\in \mathcal{D}_{ij}, \mathbf{h}_u} \left[ R_{uk_{ij}}(\mathbf{x}_u) \right]$ is the average achievable rate (taken over user location $\mathbf{x}_u$ and channel fading) when the user is located in $\mathcal{D}_{ij}$ and downloads from the cache of the $k$th nearest BS (or from the backhaul if $k = K+1$).

To obtain $\mathsf R_{uk_{ij}}$, we first derive the average rate (taken over channel fading) conditioned on given user location when $k\leq K$ as
\begin{align}
\mathbb{E}_{\mathbf{h}_u} \left[ R_{uk_{ij}}(\mathbf{x}_u) \right]   = & \mathbb{E}_{\mathbf{h}_u}
\left[ \log_2(1 + \gamma_{uk_{ij}}(\mathbf{x}_u)) \right] \nonumber \\
= & \mathbb{E}_{\mathbf{h}_u}  \left[ \log_2\left(S_{uk_{ij}} + I_{uk_{ij}} + \tfrac{\sigma^2}{P} \right) \right]  \nonumber \\
&-  \mathbb{E}_{\mathbf{h}_u}  \left[ \log_2\left(I_{uk_{ij}} + \tfrac{\sigma^2}{P} \right) \right] \label{eqn:Ehu}
\end{align}
Since $S_{uk_{ij}} + I_{uk_{ij}} = \sum_{b\in \Phi_{k_{ij}}} h_{ub} r_{ub}^{-\alpha}$ and $I_{uk_{ij}} = \sum_{b\in \Phi_{k_{ij}}, b\neq k_{ij}} h_{ub} r_{ub}^{-\alpha}$ are the sum of independent exponential distributed random variables with given $\{r_{ub}^{-\alpha}\}$, the probability density function of $S_{uk_{ij}} + I_{uk_{ij}}$ and $I_{uk_{ij}}$ with given user location can be derived, respectively, as $
f_{S_{uk_{ij}} + I_{uk_{ij}}|\mathbf{x}_u}(x) = \sum_{b\in \Phi_{k_{ij}}} \delta_{ubk_{ij}} r_{ub}^{\alpha} e^{-xr_{ub}^{\alpha}}$ and $f_{I_{uk_{ij}}|\mathbf{x}_u}(x)  = \sum_{b\in \Phi_b, b\neq k_{ij}} \delta_{ub\bar k_{ij}}r_{ub}^{\alpha} e^{-xr_{ub}^{\alpha}}$,
where $\delta_{ubk_{ij}} =\prod_{b'\in\Phi_{k_{ij}}}^{ b'\neq b} \frac{r_{ub'}^{\alpha}}{r_{ub'}^{\alpha} - r_{ub}^{\alpha}}$ and $\delta_{ub\bar k_{ij}} = \prod_{b'\in\Phi_{k_{ij}}}^{b'\neq b,k_{ij}} \frac{r_{ub'}^{\alpha}}{r_{ub'}^{\alpha} - r_{ub}^{\alpha}} $.

Then, we can derive the first term of \eqref{eqn:Ehu} as
\begin{align}
&\mathbb{E}_{\mathbf{h}_u}  \left[ \log_2\left(S_{uk_{ij}} + I_{uk_{ij}} + \tfrac{\sigma^2}{P} \right) \right] \nonumber \\
& = \sum_{b\in \Phi_{k_{ij}}} \delta_{ubk_{ij}} \int_{0}^{\infty} \log_2 \left(x + \tfrac{\sigma^2}{P}\right)r_{ub}^{\alpha} e^{-xr_{ub}^{\alpha}} {\rm d}x \nonumber\\
& = \sum_{b\in \Phi_{k_{ij}}}
\frac{\delta_{ubk_{ij}}}{\ln 2} \left(-e^{\frac{\sigma^2}{P} r_{ub}^\alpha} {\rm Ei}\left(- \tfrac{\sigma^2}{P}r_{ub}^\alpha\right) + \ln \tfrac{\sigma^2}{P}\right) \label{eqn:S+I}
\end{align}
where ${\rm Ei}(x) = -\int_{-x}^{\infty} \frac{e^{-t}}{t}dt$ denotes the exponential integral. Similar to the derivation of \eqref{eqn:S+I}, we can obtain $\mathbb{E}_{\mathbf{h}_u}  \big[ \log_2\big(I_{uk_{ij}} + \tfrac{\sigma^2}{P} \big) \big] = \sum_{b\in \Phi_{k_{ij}},b\neq k_{ij}}
\frac{\delta_{ub\bar k_{ij}}}{\ln 2} \big(-e^{\frac{\sigma^2}{P} r_{ub}^\alpha} {\rm Ei}\big(- \tfrac{\sigma^2}{P}r_{ub}^\alpha\big) + \ln \tfrac{\sigma^2}{P}\big)$. By substituting into \eqref{eqn:Ehu}, we can obtain
\begin{multline}
\mathbb{E}_{\mathbf{h}_u} \left[ R_{uk_{ij}}(\mathbf{x}_u) \right] = \delta_{uk_{ij}k_{ij}} {\sf F}_{k_{ij}}(\mathbf{x}_u )  \\+  \sum_{b\in \Phi_{k_{ij}}, b\neq k_{ij}}   (\delta_{ubk_{ij}}-\delta_{ub\bar k_{ij}}) {\sf F}_b(\mathbf{x}_u) \label{eqn:Ruk1}
\end{multline}
where ${\sf F}_b(\mathbf{x}_u) =-\frac{\exp({\frac{\sigma^2}{P} ||\mathbf{x}_u - \mathbf{y}_b||^\alpha})}{\ln 2} {\rm Ei}\big(- \frac{\sigma^2}{P}||\mathbf{x}_u - \mathbf{y}_b||^\alpha\big) + \log_2 \tfrac{\sigma^2}{P} $.

When $k = K+1$, the user associates with the nearest BS (i.e., the $i$th BS when the user is located in the $i$th cell) and downloads file from the backhaul, we derive $\mathbb{E}_{\mathbf{h}_u}\left[R_{uk_{ij}}(\mathbf x_u)\right]$ by considering $\mathbb{E}[X] = \int_0^{\infty} \mathbb{P}(X>t) {\rm d}t$ as
\begin{align}
&\mathbb{E}_{\mathbf{h}_u}\left[R_{uk_{ij}}(\mathbf x_u)\right] \nonumber \\
&= \int_{0}^{\infty} \mathbb{P}\left(\min\{W_u \log_2(1 + \gamma_{ui}(\mathbf{x}_u)) , C_u^{\rm bh}  \} > t ~|~ \mathbf{x}_u \right) {\rm d} t \nonumber\\
& = \int_{0}^{C_u^{\rm bh} } \mathbb{P}\left(W_u \log_2(1 + \gamma_{ui}(\mathbf{x}_u))  > t ~|~ \mathbf{x}_u \right) {\rm d} t  \nonumber \\
& = \int_{0}^{C_u^{\rm bh} } \mathbb{P}\left(\gamma_{ui}(\mathbf{x}_u)  > 2^{\frac{t}{W_u}} - 1~|~ \mathbf{x}_u \right) {\rm d} t \nonumber\\
& = \int_{0}^{C_u^{\rm bh}} e^{-(2^{\frac{t}{W_u}} - 1) r_{ui}^{\alpha}\frac{\sigma^2}{P}} \!\!\!\!\prod_{b\in \Phi_{i}, b \neq i} \!\!\!\left(1 + (2^{\frac{t}{W_u}} - 1) r_{ui}^{\alpha} r_{ub}^{-\alpha}\right)^{-1}\!\!  {\rm d}t \label{eqn:RuK1}
\end{align}
where  the last step is from substituting \eqref{eqn:suk}.

By averaging over user location within small region $\mathcal{D}_{ij}$, we can obtain
\begin{align}
{\sf R}_{uk} & = \mathbb{E}_{\mathbf{x}_u\in \mathcal{D}_{ij}}\left[ \mathbb{E}_{\mathbf{h}_u} \left[ R_{uk_{ij}}(\mathbf{x}_u) \right]\right] \nonumber\\
& =\frac{1}{|\mathcal{D}_{ij}|} \iint\limits_{\mathbf{x}_u \in \mathcal{D}_{ij}}\mathbb{E}_{\mathbf{h}_u} \left[ R_{uk}(\mathbf{x}_u) \right]  {\rm d}x_{u1}{\rm d}x_{u2} \label{eqn:Ruk}
\end{align}
Then, by substituting \eqref{eqn:Ruk1} and \eqref{eqn:RuK1} into \eqref{eqn:Ruk}, and finally \eqref{eqn:Ruk} into \eqref{eqn:Ru1}, Proposition 2 can be proved.

\bibliographystyle{IEEEtran}
\bibliography{dongbib}
\end{document}